\documentclass[12pt,journal]{IEEEtran}
\usepackage{epsfig,makeidx,color}
\usepackage{graphicx}
\usepackage{amsbsy}
\usepackage{amsmath}
\usepackage{amssymb}
\usepackage{euscript}
\usepackage{subfigure}
\usepackage{psfrag}
\usepackage{cite}

\newtheorem{mytheorem}{\bf Theorem}
\newtheorem{mylemma}{\bf Lemma}

\newcommand {\Define} {\stackrel {\Delta} {=}  }

\def\bh{{\boldsymbol{h}}}

\renewcommand{\baselinestretch}{1.728}
\begin{document}
\title{Single-User  Beamforming in Large-Scale MISO Systems with Per-Antenna Constant-Envelope Constraints: The Doughnut Channel}
\author{\IEEEauthorblockN{Saif Khan Mohammed* and Erik G. Larsson}
}
\onecolumn
\maketitle
\begin{abstract}
Large antenna arrays at the transmitter (TX) has recently  been  shown
to achieve remarkable intra-cell interference suppression at low complexity.
However, building large arrays in practice, would require the use of
power-efficient RF amplifiers, which generally have poor linearity characteristics
and hence would require the use of input signals with a very small
peak-to-average power ratio (PAPR). In this paper, we consider the single-user
Multiple-Input Single-Output (MISO) channel for the case where the
TX antennas are constrained to transmit signals having constant envelope (CE).
We show that, with per-antenna CE transmission the effective channel seen by the receiver is a
SISO AWGN channel with its input constrained to lie in a doughnut-shaped region.
For a broad class of fading channels,
analysis of the effective doughnut channel shows that under a per-antenna CE input constraint,
i) compared to an average-only total TX power constrained MISO channel,
the extra total TX power required to achieve a desired information rate is
small and bounded,
ii) with $N$ TX antennas an $O(N)$ array power gain is achievable, and
iii) for a desired information rate,
using power-efficient amplifiers with CE inputs would require significantly less
total TX power when compared to using highly linear (power-inefficient) amplifiers
with high PAPR inputs.
\end{abstract}
\begin{IEEEkeywords}
Beamforming, MISO, constant envelope, per-antenna.
\end{IEEEkeywords}
\IEEEpeerreviewmaketitle

{\renewcommand{\thefootnote}{} \footnote{The authors are with the Communication Systems Division, Dept. of Electrical Engineering (ISY), 
Link{\"o}ping University, Link{\"o}ping, Sweden.  This work was supported by the
    Swedish Foundation for Strategic Research (SSF) and ELLIIT.
    E. G. Larsson is a Royal Swedish Academy of Sciences (KVA)
    Research Fellow supported by a grant from the Knut and Alice
    Wallenberg Foundation. Parts of the results in this paper will be
    presented at IEEE ICC 2012.
    }}
\section{Introduction}
\vspace{-3mm}
The high electrical power consumption in cellular base stations (BS) has been recognized as a major problem worldwide \cite{Fettweis}.
One way of reducing the power consumed is to reduce the total radiated radio-frequency (RF) power.
In theory, the total radiated power from a BS can be reduced without affecting the
downlink throughput, by increasing the
number of antennas. This effect has been traditionally referred to as the ``array power gain'' \cite{Tse-book}.
In addition to improving power-efficiency, there has been a great deal of recent interest in
multi-user Multiple-Input Multiple-Output (MIMO) systems with {\em large antenna arrays} \cite{SPM-paper,Marzetta},
due to their ability to substantially reduce intra-cell interference with very simple
signal processing. 
In general, multiple antenna beamforming is a well known technology to improve link performance \cite{Zhang}.

\begin{figure}[t]
\centering
\subfigure[ MRT]{
\psfrag{u}[][][0.9][0]{$u$}
\psfrag{h_1}[][][0.9][0]{$\sqrt{P_T} \frac{h_1^*}{\Vert \bh \Vert}$}
\psfrag{h_i}[][][0.9][0]{$\sqrt{P_T} \frac{h_i^*}{\Vert \bh \Vert}$}
\psfrag{h_N}[][][0.9][0]{$\sqrt{P_T} \frac{h_N^*}{\Vert \bh \Vert}$}
\psfrag{Dynamic range}[][][0.6][0]{$\mbox{Amplitude range} \,=\, [ 0  \cdots \sqrt{P_T} \vert u \vert] $}
\psfrag{e_j1}[][][0.9][0]{$e^{j \theta_1^u}$}
\psfrag{e_ji}[][][0.9][0]{$e^{j \theta_i^u}$}
\psfrag{e_jN}[][][0.9][0]{$e^{j \theta_N^u}$}
\psfrag{Dynamic range2}[][][0.6][0]{$\mbox{\textbf{Constant} Amplitude} \,\,= \sqrt{\frac{P_T}{N}} $}
\psfrag{g_1}[][][0.9][0]{$\sqrt{\frac{P_T}{N}}$}
\psfrag{g_i}[][][0.9][0]{$\sqrt{\frac{P_T}{N}}$}
\psfrag{g_N}[][][0.9][0]{$\sqrt{\frac{P_T}{N}}$}
\includegraphics[width=0.32\textwidth]{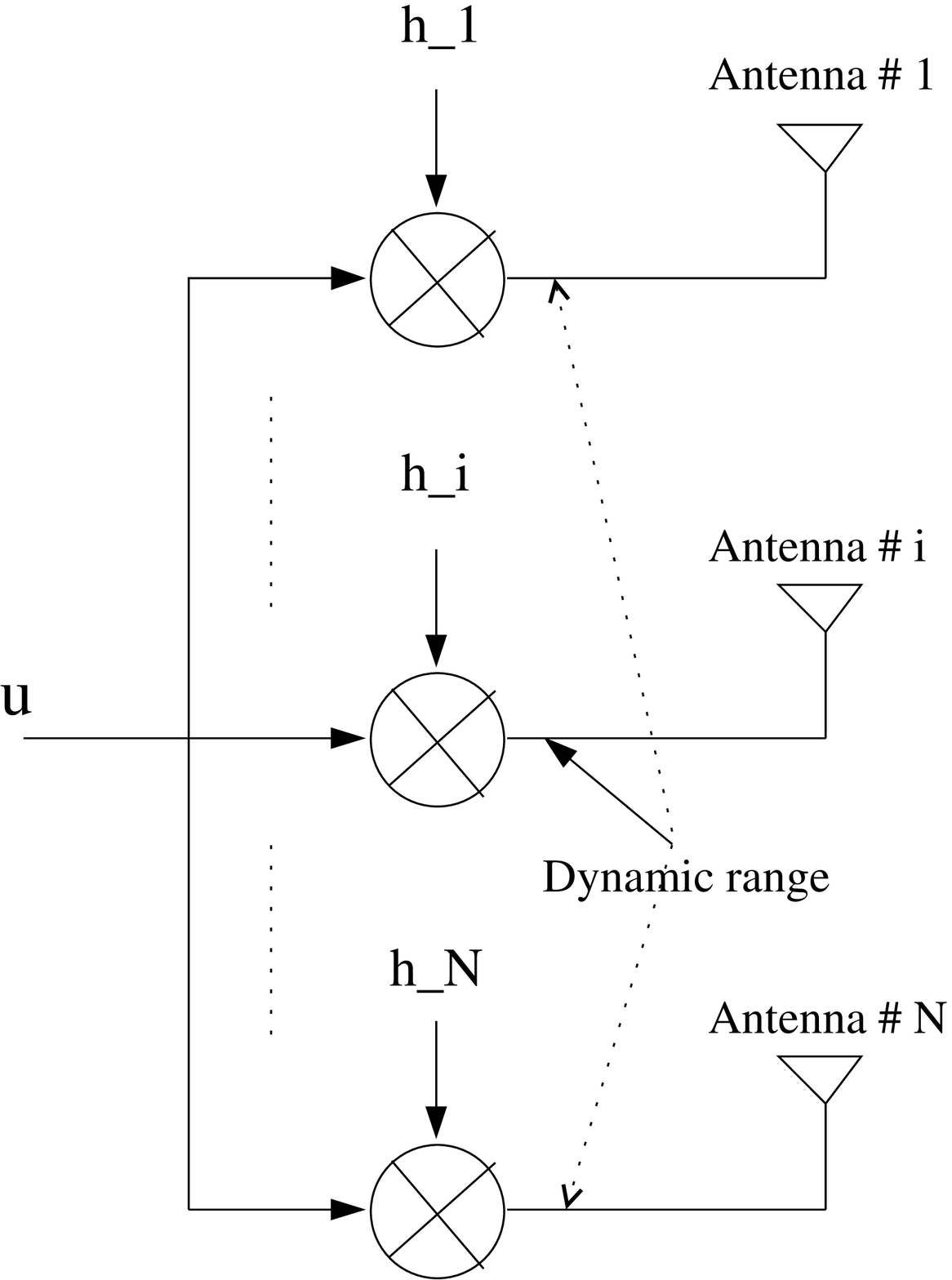}
\label{fig:intr_a}
}
\subfigure[ CE ]{
\psfrag{u}[][][0.9][0]{$u$}
\psfrag{h_1}[][][0.9][0]{$\sqrt{P_T} \frac{h_1^*}{\Vert \bh \Vert}$}
\psfrag{h_i}[][][0.9][0]{$\sqrt{P_T} \frac{h_i^*}{\Vert \bh \Vert}$}
\psfrag{h_N}[][][0.9][0]{$\sqrt{P_T} \frac{h_N^*}{\Vert \bh \Vert}$}
\psfrag{Dynamic range}[][][0.6][0]{$\mbox{Amplitude range} \,=\, [ 0  \cdots \sqrt{P_T} \vert u \vert] $}
\psfrag{e_j1}[][][0.9][0]{$e^{j \theta_1^u}$}
\psfrag{e_ji}[][][0.9][0]{$e^{j \theta_i^u}$}
\psfrag{e_jN}[][][0.9][0]{$e^{j \theta_N^u}$}
\psfrag{Dynamic range2}[][][0.6][0]{$\mbox{\textbf{Constant} Amplitude} \,\,= \sqrt{\frac{P_T}{N}} $}
\psfrag{g_1}[][][0.9][0]{$\sqrt{\frac{P_T}{N}}$}
\psfrag{g_i}[][][0.9][0]{$\sqrt{\frac{P_T}{N}}$}
\psfrag{g_N}[][][0.9][0]{$\sqrt{\frac{P_T}{N}}$}
\includegraphics[width=0.29\textwidth]{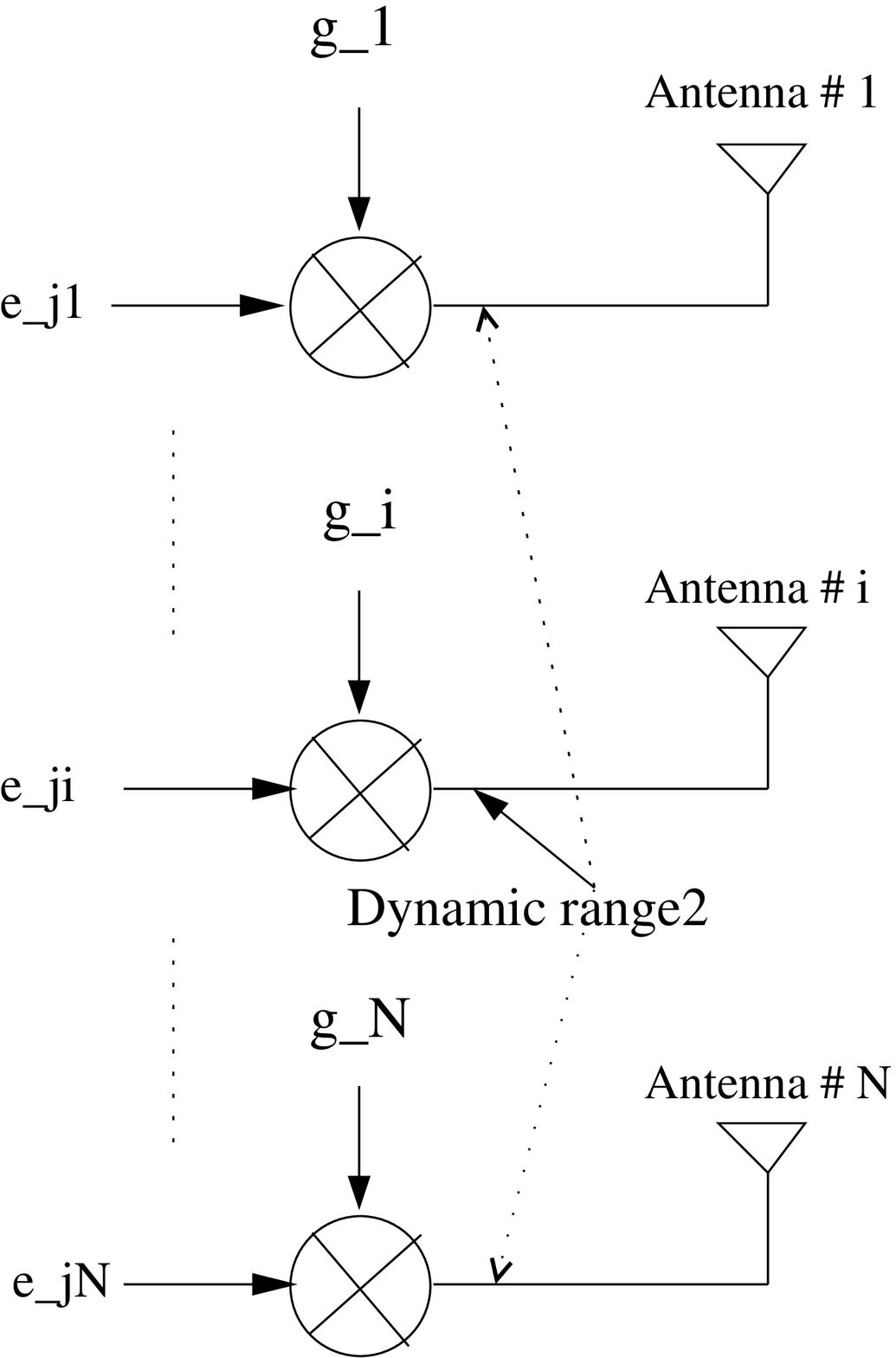}
\label{fig:intr_b}
}
\caption{Maximum Ratio Transmission (MRT) versus per-antenna Constant Envelope (CE) constrained transmission,
for a given average total transmit power constraint of $P_T$. ${\bf h}=(h_1,\cdots,h_N)^T$ is the vector of complex channel gains.}
\vspace{-7mm}
\end{figure}

To illustrate the improvement in power efficiency with large antenna arrays, let us consider a MISO channel between a transmitter (TX) having $N > 1$ antennas and a single-antenna receiver.
With knowledge of the channel vector (${\bf h} =(h_1,h_2,\cdots,h_N)^T$) at the TX and
an {\em average-only} total transmit power constraint of $P_T$,
an information symbol $u$ (with mean energy ${\mathbb E}[\vert u \vert^2] = 1$) can be beamformed in such a way (the $i$-th antenna transmits $\sqrt{P_T} h_i^* u / \Vert {\bf h} \Vert_{_2}$)
that the signals from different TX antennas add up {\em coherently}
at the receiver (the received signal is $\sqrt{P_T} \Vert {\bf h} \Vert_{_2} u$), thereby resulting in an effective channel
with a received signal power that is $\Vert {\bf h} \Vert_{_2}^2 / \vert h_1 \vert^2$ times higher compared to a scenario where the TX has
only one antenna. For a broad class of fading channels (e.g., i.i.d. fading, single-path
direct-line-of-sight (DLOS)) $\Vert {\bf h} \Vert_{_2}^2 = \vert h_1 \vert^2 O(N)$, and
  therefore, for a fixed desired received signal power,
the total transmit power can be reduced by roughly half
with every doubling of the
number of TX antennas.
This type of beamforming is referred to as ``Maximum Ratio Transmission'' (MRT)
(see Fig.~\ref{fig:intr_a}).

In theory, to achieve an order of magnitude reduction in the total radiated power (without affecting throughput) we would need TX with a
large  number of antennas (by large, we mean tens or even hundreds \cite{SPM-paper,Marzetta}).
However, building very large arrays in practice requires that each individual
antenna, and its associated RF electronics, be  cheaply
manufactured  and implemented in  power-efficient technology.
It is known that conventional BS are {\em highly power-inefficient}. Typically, the ratio of radiated
power to the total power consumed is less than $5$ percent, the main reason being the use of highly
{\em linear} and power-inefficient analog devices like the power amplifier \cite{Mancuso}.\footnote{In
a conventional BS, about $40-50$ percent of the total operational power is consumed by the power amplifier
and the associated RF electronics, which have a  power efficiency of only about $5-10$ percent \cite{Mancuso}.}
Generally, high linearity implies low power efficiency and vice-versa.
Therefore, non-linear  but highly power-efficient
amplifiers must be used.
With non-linear power amplifiers, the signal transmitted from each antenna must
have a {\em low  peak-to-average-power-ratio}, so as to avoid significant signal distortion.
The type of signal that
facilitates the use of the most power-efficient and cheap power amplifiers/analog components is therefore a {\em
constant envelope} (CE) signal.

With this motivation, in this paper, we consider a single-user Gaussian MISO fading channel with
the signal transmitted from each TX antenna constrained to have a constant envelope.
Fig.~\ref{fig:intr_b} illustrates the proposed signal transmission under a per-antenna CE constraint.
Essentially, for a given information symbol $u$ to be communicated to the single-antenna receiver,
the signal transmitted from the $i$-th antenna is $\sqrt{P_T/N} e^{j \theta_i^u}$.
The transmitted phase angles $(\theta_1^u, \cdots, \theta_N^u)$ are determined in such a way that
the noise-free signal received matches closely  with $u$.
The amplitude of the signal transmitted from each antenna is {\em constant} and equal to  $\sqrt{P_T/N}$
for every channel-use, irrespective of the channel realization.
By way of  contrast, with MRT, the amplitude of the transmitted signal {\em depends upon the channel realization} as well as on $u$, and can vary from $0$
to $\sqrt{P_T} \vert u\vert$.
Since  the CE constraint is much more {\em restrictive} than the average-only total power constraint in MRT,
a natural question which arises now  is   how much array power gain can be achieved with the  per-antenna
CE constraint.
Also, compared to MRT, how much extra total transmit power  is required with   per-antenna CE transmission  
to achieve a given information rate? 

So far, in the open literature, these questions have not been addressed.
For the special case of $N=1$ (SISO AWGN), the channel capacity
under a CE input constraint has been reported in \cite{Wyner}.
{
However for $N > 1$, known reported works on per-antenna power constrained communication consider an average-only or peak-only power
constraint \cite{Palomar, Mai, Durisi, WeiYu, Kemal, Shi}.
For the single-user scenario, in \cite{Palomar}, the author considers the problem of finding the optimal transmit and receive matrices which maximize the received
signal-to-noise-and-interference-ratio (SINR) in a MIMO channel, subject to a per-antenna average power constraint at the TX.
In \cite{Mai}, the author has derived a closed-form expression for the capacity of a single-user MISO channel with a per-antenna average
power constraint at the TX.
In \cite{Durisi}, the authors compute bounds on the capacity of a noncoherent single-user MIMO channel with peak per-antenna power constraints
at the TX.
}

{
For the multiuser MIMO broadcast channel with per-antenna power constraints, in \cite{WeiYu} the authors consider
minimization of the per-antenna average power radiated by the transmitter subject to a minimum SINR constraint for each user
in the downlink. They propose efficient numerical methods for solving this problem using uplink-downlink duality.
In \cite{Kemal}, the authors study the optimal multi-user linear zero-forcing beamformer which maximizes the minimum information rate to the downlink users,
under per-antenna average power constraints at the BS.
In \cite{Shi}, the authors
consider the scenario where users can also have multiple antennas, and propose methods to maximize the weighted sum-rate and max-min rate
under a per-antenna average power constraint at the BS.
}

In contrast to the above works on per-antenna {\it average/peak} power constrained communication,
in this paper, we consider the more   stringent per-antenna constant-envelope constraint where each antenna emits a signal of {\it constant} amplitude $\sqrt{ P_T/N}$.

The specific contributions presented in this paper are:
i) we show that, under a per-antenna CE constraint at the TX, the MISO channel reduces to a SISO AWGN channel with
   the noise-free received signal being constrained to lie in a ``doughnut'' shaped region in the complex plane,
ii) using the equivalent doughnut channel model, we derive analytical upper and lower bounds on the MISO channel capacity
under per-antenna CE transmission,
iii) under per-antenna CE transmission, for large $N$ we show that the optimal information alphabet (in terms of achieving
capacity) is discrete-in-amplitude and uniform-in-phase, and
iv) we also propose   novel  algorithms for transmit precoding under the per-antenna CE constraint.
Our analysis shows that for a large class of fading channels (i.i.d. Rayleigh fading, i.i.d. fading channels where the channel gains are bounded\footnote{\footnotesize{In practice, real-world channels generally have bounded channel gains.}}, DLOS),
i) under the per-antenna CE constraint, an array power gain of $O(N)$ is  indeed achievable  with $N$ antennas,
ii) for a desired information rate to be achieved,
 compared to the MRT precoder with an average-only total transmit power
constraint, the extra total transmit power required under per-antenna CE transmission is small and bounded,
iii) by using a sufficiently large antenna array, at high total transmit power $P_T$, the ratio of
the information rate achieved under the per-antenna CE constraint to the capacity of the average-only total transmit power constrained
MISO channel can be guaranteed to be {\em close to $1$}, with high probability.
This stands in   contrast  to Wyner's result in \cite{Wyner} for $N=1$, where this ratio is {\em only $1/2$} at high
$P_T$.
Analytical results are supported with numerical results for the i.i.d.\ Rayleigh fading channel.
The analysis and algorithms presented are general and   applicable to systems with any number of transmit antennas. 

\textbf{Notation:}
${\mathbb C}$ and ${\mathbb R}$ denote the set of complex and real numbers.
$\vert x \vert$, $x^*$ and $\arg(x)$ denote the absolute value, complex conjugate and argument of $x \in {\mathbb C}$ respectively.
For any positive $p \geq 1$, $\Vert {\bf h} \Vert_p \Define (\sum_i \vert h_i \vert^p)^{1/p}$ denotes the Euclidean $p$-norm of ${\bf h}=(h_1,\cdots,h_N) \in {\mathbb C}^{N}$.
${\mathbb E}[\cdot]$ denotes the expectation operator.
$\log(\cdot)$ denotes the natural logarithm, and $\log_2(\cdot)$ denotes the base-$2$ logarithm.
Abbreviations: r.v. (random variable), bpcu (bits-per-channel-use), p.d.f. (probability density function).
\section{System model}\label{Sysmodel}
We consider a single-user MISO system.
The complex channel gain between the $i$-th transmit antenna and the
single antenna receiver is denoted by $h_{i}$, and the total channel vector is denoted by
${\bf h} = ( h_{1} , h_{2} , \cdots, h_{N} )^T$.
TX is assumed to have perfect knowledge\footnote{\footnotesize{For large $N$, with Time-Division-Duplex (TDD)
communication and assuming a reciprocal channel, channel measurements at the TX using reverse link pilot signals can be used
to estimate the forward channel. {A preliminary study done by us reveals that, the performance of the proposed
CE transmission scheme degrades with increasing estimation error variance. However, interestingly, with i.i.d. Rayleigh fading the
performance loss is small even when the standard deviation of the estimation error is of the same order as the average channel gain.}}}of ${\bf h}$, whereas
the receiver is required to have only partial knowledge (we shall discuss this later in more detail).
Let the complex symbol transmitted from the $i$-th antenna be denoted by $x_i$.
The complex symbol received is 
{
\vspace{-3mm}
\begin{equation}
\label{sysmodel}
{y} = \sum_{i=1}^N h_i x_i + { w}
\end{equation}
}
where
${ w}$
denotes the circularly symmetric AWGN having mean zero and
variance $\sigma^2$, i.e., ${\mathcal C}{\mathcal N}(0,\sigma^2)$.
Due to the CE constraint on each antenna and assuming a total transmit power constraint of $P_T$,
we must have $\vert x_i \vert^2 = P_T/N \,,\,i=1,\ldots,N$.
Therefore $x_i$ must be of the form
\begin{equation}
\label{equi_power_theta}
x_i = \sqrt{\frac{P_T}{N}} e^{j \theta_i}\,\,,\,\,i=1,2,\ldots,N
\end{equation}
where $j \Define \sqrt{-1}$, and $\theta_i \in [ -\pi \,,\, \pi)$ is
the phase of $x_i$. We refer to the type of signal transmission in (\ref{equi_power_theta})
as ``CE transmission''.
Note that under an average-only total transmit power constraint, the transmitted signals are {\em only}
required to satisfy ${\mathbb E}[ \sum_i \vert x_i \vert^2] = P_T$, which is much  less restrictive  than (\ref{equi_power_theta}).
Under CE transmission, the signal received is given by (using (\ref{sysmodel}) and (\ref{equi_power_theta}))
\begin{equation}\label{eq:yk}
y=\sqrt{\frac{P_T}{N}} \sum_{i=1}^N h_{i} e^{j\theta_i} + w.
\end{equation}
Let  ${\Theta} \Define
(\theta_1, \theta_2, \cdots, \theta_N)^T$ denote the vector of
transmitted phase angles and let 
 ${ u}
\in {\mathcal U} \subset {\mathbb C}$ denote the information symbol to be communicated
to the receiver, where ${\mathcal U}$ is the information symbol alphabet.
For a given ${u}$, the precoder in the transmitter uses a map
$\Phi(\cdot) : {\mathcal U}  \rightarrow [-\pi,\pi)^N$ to generate the
transmit phase angle vector, i.e., $ 
\Theta = \Phi(u)$.
Let the set of possible
noise-free received signals scaled down by $\sqrt{P_T}$, i.e., $\sqrt{\frac{1}{N}} \sum_{i=1}^N h_{i} e^{j\theta_i}$,
be given by
{
\vspace{-6mm}
\begin{eqnarray}
\label{Mh_def}
\hspace{20mm} {\mathcal M}({\bf h}) \Define {\Big \{}  \frac{\sum_{i=1}^N h_{i} e^{j\theta_i}}{\sqrt{N}}  \,\,,\,\, \theta_i \in [-\pi,\pi) \,\,i=1,\ldots,N {\Big \}}
\end{eqnarray}
}
By choosing ${\mathcal U} \subseteq {\mathcal M}({\bf h})$,
for any $u \in {\mathcal U}$, it is implied that
$u \in {\mathcal M}({\bf h})$, and therefore from (\ref{Mh_def}) it follows that, there exists a phase angle vector $\Theta^u = (\theta_1^u,\cdots, \theta_N^u)$
such that\footnote{\footnotesize{${\mathcal U} \subseteq {\mathcal M}({\bf h})$
implies that the information symbol alphabet must be chosen adaptively with ${\bf h}$.  Therefore the receiver must be informed
about the newly chosen ${\mathcal U}$, every time it changes. However,   we shall see in Section~\ref{set_study} that
the set ${\mathcal M}({\bf h})$ is the interior of a ``doughnut'' shaped region in the 2-dimensional complex-plane and can therefore be fully characterized
with only two non-negative real numbers (the inner and the outer radius of the doughnut). Hence, the TX only needs to inform the receiver about these two numbers every time ${\bf h}$ changes.
}}
{
\vspace{-1mm}
\begin{equation}
\label{u_sat}
u = \sqrt{\frac{1}{N}} \sum_{i=1}^N h_{i} e^{j\theta_i^u}.
\end{equation}
}
With the precoder map
{
\vspace{-6mm}
\begin{eqnarray}
\label{prec_def}
\Phi(u) \Define \Theta^u
\end{eqnarray}
}
where $\Theta^u$ satisfies (\ref{u_sat}), the received signal is given by
{
\vspace{-4mm}
\begin{eqnarray}
\label{doughnut_eqn}
y = \sqrt{P_T}  \, u + w
\end{eqnarray}
}
{\em i.e., the noise-free received signal is the same as the intended information symbol $u$ scaled up by $\sqrt{P_T}$.}
Subsequently in this paper,
we propose to choose ${\mathcal U} \subseteq {\mathcal M}({\bf h})$,\footnote{\footnotesize{For ${\mathcal U}
\not \subseteq {\mathcal M}({\bf h})$, it may be possible
to consider a precoder map which satisfies (\ref{u_sat}) for $u \in {\mathcal M}({\bf h})$, and for any $u \notin {\mathcal M}({\bf h})$ finds the phase angle vector which
minimizes the non-zero energy of the residual/error term ${\Big (} \frac{\sum_{i=1}^N h_{i} e^{j\theta_i}}{\sqrt{N}} - u {\Big )}$. However, even with such an error-minimizing precoder,
it has been observed via simulations that
having ${\mathcal U} \not \subseteq {\mathcal M}({\bf h})$  does not increase the achievable information rate compared to when ${\mathcal U} \subseteq {\mathcal M}({\bf h})$.}}
and define the  precoder map  as  in (\ref{prec_def}) and (\ref{u_sat}).
With ${\mathcal U} \subseteq {\mathcal M}({\bf h})$ it is clear that the information rate depends  on ${\mathcal M}({\bf h})$.
In the next section, we give a more detailed characterization of ${\mathcal M}({\bf h})$.
{
\vspace{-1mm}
}
\section{Characterization of ${\mathcal M}({\bf h})$}\label{set_study}
We characterize ${\mathcal M}({\bf h})$ through a series of intermediate results.
First, we define the maximum and minimum absolute value of any complex
number in ${\mathcal M}({\bf h})$.
\begin{eqnarray}
\label{max_min_def}
M({\bf h})  \Define  \max_{\Theta = (\theta_1,\cdots,\theta_N)\,,\,\theta_i \in [-\pi,\pi)} {\Bigg \vert }\frac{\sum_{i=1}^N h_i e^{j \theta_i}}{\sqrt{N}} {\Bigg \vert} \,\,,\,\,
m({\bf h})  \Define  \min_{\Theta = (\theta_1,\cdots,\theta_N)\,,\,\theta_i \in [-\pi,\pi)} {\Bigg \vert }\frac{\sum_{i=1}^N h_i e^{j \theta_i}}{\sqrt{N}} {\Bigg \vert}
\end{eqnarray}
\begin{mylemma}\label{su_circularsymmetry}
If $z \in {\mathcal M}({\bf h})$ then so does
$z e^{j \phi}$ for all $\phi \in [-\pi,\pi)$.
\end{mylemma}
{\it Proof} --
Since $z \in {\mathcal M}({\bf h})$, from (\ref{Mh_def}) it follows that there exists a phase
vector $\Theta^z = (\theta_1^z,\theta_2^z,\cdots,\theta_N^z)$ such that
$z = \frac{\sum_{i=1}^N h_i e^{j \theta_i^z}} {\sqrt{N}}$.
Consider the phase angle vector
${\Tilde \Theta^z} = ({\Tilde \theta_1^z},{\Tilde \theta_2^z},\cdots,{\Tilde \theta_N^z})$
with
${\Tilde \theta_i^z} = \theta_i^z + \phi \,,\,i=1,2,\cdots,N$.
It now follows that,
$z e^{j \phi} =   e^{j \phi} \frac{\sum_{i=1}^N  h_i e^{j \theta_i^z}} {\sqrt{N}}  = \frac {\sum_{i=1}^N h_i e^{j (\theta_i^z + \phi)}} {\sqrt{N}}  =  \frac {\sum_{i=1}^N h_i e^{j {\Tilde \theta_i^z}}} {\sqrt{N}} \in {\mathcal M}({\bf h})$.
$\hfill\blacksquare$

Essentially Lemma \ref{su_circularsymmetry} shows that the set ${\mathcal M}({\bf h})$
exhibits a circular symmetry in ${\mathbb C}$.
The following two lemmas characterize $M({\bf h})$ and $m({\bf h})$.
\begin{mylemma}\label{max_val}
{
\vspace{-3mm}
\begin{eqnarray}
M({\bf h}) = \frac { \sum_{i=1}^N \vert h_i \vert } {\sqrt{N}} = \frac {\Vert {\bf h} \Vert_1} {\sqrt{N}} .
\end{eqnarray}
}
\vspace{-4mm}
\end{mylemma}
{\it Proof} --
The proof essentially follows from the extended triangular inequality
${\Bigg \vert } \sum_{i=1}^N   h_i e^{j \theta_i} {\Bigg \vert }   \leq \sum_{i=1}^N \vert h_i \vert
$ with equality achieved when $e^{j \theta_i} = \frac{{h_i^*}}{\vert h_i \vert}$, i.e., $\theta_i = - \arg(h_i)$.
$\hfill\blacksquare$

{
\vspace{-3mm}
\begin{mylemma}\label{min_val_bnd}
\begin{eqnarray}
\label{mh_upp_bnd}
m({\bf h}) \leq \frac{\Vert {\bf h} \Vert_{\infty}}{\sqrt{N}} = \frac {\max_{i=1,\ldots,N} \vert h_i \vert}{ \sqrt{N}}.
\end{eqnarray}
\end{mylemma}
}
{\it Proof} --
Let the absolute values of the components of ${\bf h}$ be ordered as
$\vert h_{i_1} \vert \geq \vert h_{i_2} \vert \geq \ldots \geq \vert h_{i_N} \vert$.
By choosing the phase angles to be
$\theta_{i_k}  =  - \arg(h_{i_k})$ for odd $k$ and
$\theta_{i_k}  = - (\arg(h_{i_k}) + \pi) $ for even $k$,
for even $N$ we have
$\sum_{i=1}^N  h_i e^{j \theta_i}  =   \sum_{k=1}^{N} h_{i_k} e^{j \theta_{i_k}} = \sum_{k=1}^{N/2} {\big (} \vert h_{i_{2k -1}} \vert - \vert h_{i_{2k}} \vert {\big )}$
$ \leq  \sum_{k=1}^{N-1} {\big (} \vert h_{i_k} \vert - \vert h_{i_{k+1}} \vert {\big )} = \vert h_{i_1} \vert  -  \vert h_{i_N} \vert   \leq  \vert h_{i_1} \vert = \Vert {\bf h} \Vert_{\infty}$.
Similarly, for odd $N$, we have
$\sum_{i=1}^N   h_i e^{j \theta_i}  =  {\Big [} \sum_{k=1}^{(N-1)/2}  {\big (} \vert h_{i_{2k -1}} \vert - \vert h_{i_{2k}} \vert {\big )} {\Big ]} + \vert h_{i_N} \vert $
$ \leq   {\Big [}  \sum_{k=1}^{N-2} {\big (} \vert h_{i_k} \vert - \vert h_{i_{k+1}} \vert  {\big )}  {\Big ]}  +  \vert h_{i_N} \vert 
 =  \vert h_{i_1} \vert -  \vert h_{i_{N-1}} \vert + \vert h_{i_N} \vert 
 \leq  \vert h_{i_1} \vert = \Vert {\bf h} \Vert_{\infty}$.
The proof now follows from the definition of $m({\bf h})$.
$\hfill\blacksquare$

\begin{figure}[t]
\begin{center}
\hspace{-4mm}
\epsfig{file=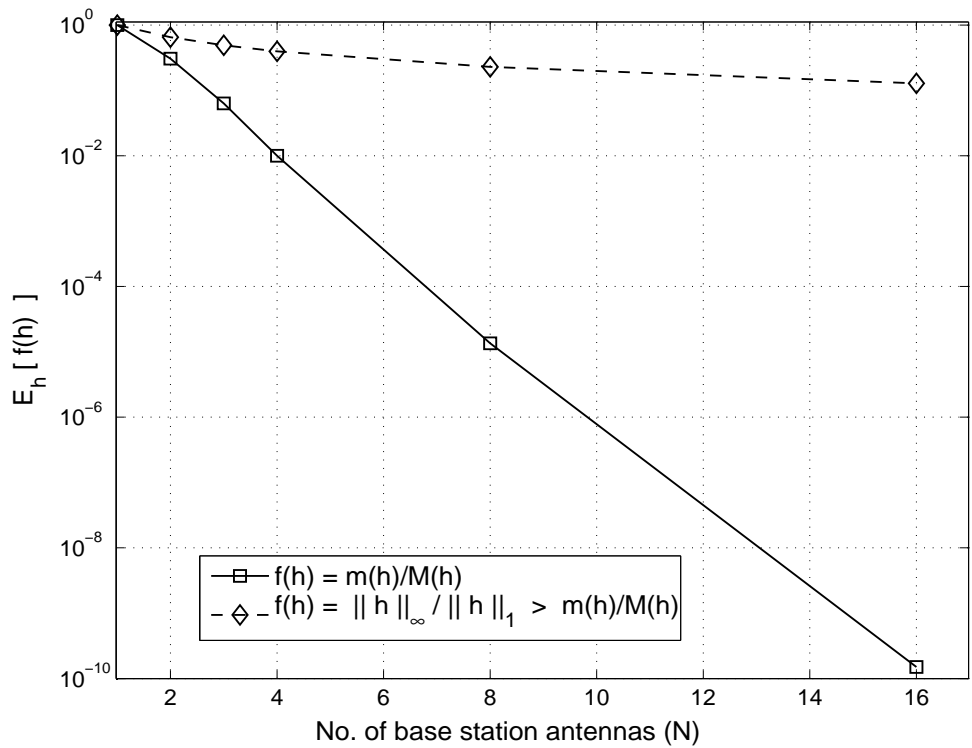, width=130mm,height=100mm}
\end{center}
\vspace{-4mm}
\caption{
Mean value of the ratio $m({\bf h})/M({\bf h})$ as a function of increasing $N$, for i.i.d.\ ${\mathcal C}{\mathcal N}(0,1)$ Rayleigh fading.
We calculate $m({\bf h})$ in (\ref{max_min_def}) using an iterative gradient descent type method.
}
\label{fig_mh_Mh_ratio}
\vspace{-4mm}
\end{figure}

{
In Appendix \ref{app_mh}, for the i.i.d. Rayleigh fading channel, we analytically show that for any constant $c > 0$,
$\lim_{N \rightarrow \infty} \mbox{\footnotesize{Prob}}{\Big (} m({\bf h}) \, \geq \, \frac { c \log(N)}{\sqrt{N}}   {\Big )}  \,\, = \,\, 0$,
which essentially means that for any arbitrarily small $\epsilon > 0$, there exists a corresponding integer $N(\epsilon,c)$ such that
$\mbox{\footnotesize{Prob}}{\Big (} m({\bf h}) \, \geq \, \frac { c \log(N)}{\sqrt{N}}   {\Big )} \leq \epsilon$ for all $N \geq N(\epsilon,c)$.
Basically, it means that for sufficiently large $N$, with very high probability $m({\bf h}) \leq \frac { c \log(N)}{\sqrt{N}}$.
Since $\frac { c \log(N)}{\sqrt{N}} \rightarrow 0$ as $N \rightarrow \infty$, it follows that with increasing $N$,  $m({\bf h})$ approaches $0$
with high probability.
A similar result has been stated in \cite{Hassibi}, where it has been shown that for large $N$, $\Vert {\bf h} \Vert_{\infty} = \max_i \vert h_i \vert = {\mathbb E}[\vert h_i \vert] O(\log(N))$.
}
Numerical results for the i.i.d.\ Rayleigh fading channel have however revealed that, with increasing $N$, $m({\bf h})$
goes to zero at a significantly faster rate than $\log(N)/\sqrt{N}$ (implying that the upper bound in (\ref{mh_upp_bnd}) is not quite tight).
We illustrate this fact in Fig.~\ref{fig_mh_Mh_ratio}, where we plot the mean value of
the ratio $m({\bf h})/M({\bf h})$ and its upper bound $\Vert {\bf h} \Vert_{\infty} / \Vert {\bf h} \Vert_1$ as
a function of increasing $N$.
{
For i.i.d. fading channels where the channel gains are bounded, i.e., $\vert h_i \vert \leq M \,\, \forall i=1,2,\cdots,N$ for some constant $M$,
it follows that $\Vert {\bf h} \Vert_{\infty}$ is also bounded ($\Vert {\bf h} \Vert_{\infty} \leq M$) and hence
$\Vert {\bf h} \Vert_{\infty}\, / \, \sqrt{N}$ will converge to $0$ as $N \rightarrow \infty$.
Since, $m({\bf h}) \leq \Vert {\bf h} \Vert_{\infty}\, / \, \sqrt{N}$, it immediately follows that $m({\bf h}) \rightarrow 0$ as $N \rightarrow \infty$.\footnote{\footnotesize{We conjecture that $m({\bf h})  \rightarrow 0$ as $N \rightarrow \infty$ even if the i.i.d. channel gain distribution has {\it unbounded}
support, though we do not have a rigorous proof of this statement.}}
}
For the single-path only DLOS channel with $\vert h_1 \vert = \cdots = \vert h_N \vert$, it can be shown that for any $N \geq 2$ and any ${\bf h}$,
$m({\bf h}) = 0$. (With $\theta_i = \frac{ 2\pi (i-1) }{N} - \arg(h_i), i=1,2,\ldots,N$, it is clear that $\sum_i h_i e^{j \theta_i} = 0$.)

The next theorem characterizes the set ${\mathcal M}({\bf h})$.

\begin{mytheorem}\label{set_Mh}
\begin{equation}
\label{set_Mh_ch}
{\mathcal M}({\bf h}) = {\Big \{} z \,\,\vert \,\, z \in {\mathbb C}\,,\, m({\bf h}) \leq \vert z \vert \leq M({\bf h}) {\Big \}}.
\end{equation}
\end{mytheorem}

{\it Proof} --
Let
{
\vspace{-4mm}
\begin{equation}
\label{thetastar_def}
(\theta_1^\star,\theta_2^\star,\cdots,\theta_N^\star) \Define \arg \hspace{-6mm} \min_{\theta_i \in [-\pi,\pi)\,,\,i=1,2,\ldots,N} {\Bigg \vert}  \frac { \sum_{i=1}^N h_i e^{j \theta_i} } {\sqrt{N}} {\Bigg \vert }
\end{equation}
}
Consider the single variable function
{
\vspace{-4mm}
\begin{eqnarray}
\label{ft_def}
f(t) \Define {\Bigg \vert } \frac { \sum_{i=1}^N h_i e^{j \theta_i(t)} } {\sqrt{N}}   {\Bigg \vert }^2  \,\,,\,\, t \in [0,1]
\end{eqnarray}
}
where the functions $\theta_i(t)\,,\,i=1,2,\ldots,N$ are defined as
{
\vspace{-4mm}
\begin{equation}
\label{theta_it}
\theta_i(t) \Define (1 -t) \theta_i^\star  - t \arg(h_i) \,\,,\,\, t \in [0,1].
\end{equation}
}
Note that $f(t)$ is a differentiable function of $t$, and therefore it is continuous for all $t \in [0,1]$.
Also from (\ref{thetastar_def}), Lemma \ref{max_val} and (\ref{max_min_def}) it follows that
{
\vspace{-6mm}
\begin{eqnarray}
f(0)  =  m({\bf h})^2 \,\,\,,\,\,\,
f(1)  =  M({\bf h})^2
\end{eqnarray}
}
Since $f(t)$ is continuous, it follows that for any non-negative real number $c$ with $m({\bf h})^2 \leq c^2 \leq M({\bf h})^2$, there
exists a value of $t=t^\prime \in [0 ,1]$ such that
{
\vspace{-6mm}
\begin{equation}
\label{ftstar}
f(t^\prime) = c^2.
\end{equation}
}
Let
{
\vspace{-7mm}
\begin{equation}
\label{zstar_def}
z^\prime \Define \frac { \sum_{i=1}^N  h_i  e^{j \theta_i(t^\prime)} } {\sqrt{N}}.
\end{equation}
}
From the definition of ${\mathcal M}({\bf h})$ in (\ref{Mh_def}) it is clear that
$z^\prime \in {\mathcal M}({\bf h})$.
From (\ref{ftstar}), (\ref{zstar_def}) and (\ref{ft_def})  it follows that
\begin{equation}
\vert z^\prime \vert = \sqrt{ f(t^\prime)} = c.
\end{equation}
Therefore, we have shown that for any non-negative real number $c \in [ m({\bf h}) \,,\, M({\bf h})]$, there exists
a complex number having modulus $c$ and belonging to ${\mathcal M}({\bf h})$.

Further, from Lemma \ref{su_circularsymmetry}, we already know that the set ${\mathcal M}({\bf h})$ is circularly symmetric, and
therefore all complex numbers with modulus $c$ belong to ${\mathcal M}({\bf h})$.
Since the choice of $c \in [ m({\bf h}) \,,\, M({\bf h})]$ was arbitrary,
any complex number with modulus in the interval $[ m({\bf h}) \,,\, M({\bf h})]$ belongs to ${\mathcal M}({\bf h})$.
$\hfill\blacksquare$
\subsection{The proposed precoder map $\Phi(u) = \Theta^u$}
\label{ce_prec}
The proof of Theorem \ref{set_Mh} is constructive and for a given $u \in {\mathcal U} \subseteq {\mathcal M}({\bf h})$,
it gives us a method to find the corresponding phase angle vector $\Theta^u = (\theta_1^u,\cdots,\theta_N^u)$ which satisfies (\ref{u_sat}).
For a given $u \in {\mathcal U} \subseteq {\mathcal M}({\bf h})$, we define the function
$f_u(t) \Define f(t) - \vert u \vert^2 \,\,,\,\, t \in [0,1]$
where $f(t)$ is given by (\ref{ft_def}).
Using Newton-type methods or simple brute-force enumeration,
we can find a $t = t_u$ satisfying $f_u(t_u) = 0$ (the existence of such a $t_u$ is guaranteed by the constructive proof of Theorem \ref{set_Mh}).
The phase angles which satisfy
(\ref{u_sat}) are then given by
$\theta_i^u = \theta_i(t_u) + \phi$
where $\theta_i(t)$ is given by (\ref{theta_it}), and $\phi$ is given by
$e^{j \phi} = \frac { u \sqrt{N}} { \sum_{i=1}^N  h_i  e^{j \theta_i(t_u)} }$.

Yet another method to obtain $\Theta^u$  is to minimize the  error norm function $e^u(\Theta) \Define \vert u - \sum_{i=1}^N h_i e^{j \theta_i} /\sqrt{N} \vert^2$ w.r.t.\ $\Theta$.
For large $N$, it has been observed that,  most local minima of the error norm function
have small error norms, and therefore  low-complexity  methods like
 gradient descent   can be used.\footnote{\footnotesize{{One method, that we have empirically found to have fast local minima convergence, is to sequentially update one phase angle at a time while keeping the others fixed
in such a way that the objective function value $e^u(\Theta) \Define \vert u - \sum_{i=1}^N h_i e^{j \theta_i} /\sqrt{N} \vert^2$ decreases with every update.
Each update is a simple one-dimensional optimization problem, and since the convergence is fast, the order of complexity is expected to be the same as the MRT scheme, i.e., $O(N)$.}}}
For very small $N=2,3$ there exist closed-form expressions for $\Theta^u$ as shown below.\footnote{\footnotesize{
{
For $N=2$, $m({\bf h})= {\big \vert} \vert h_1 \vert \, - \, \vert h_2 \vert {\big \vert} / \sqrt{2}$, and
$M({\bf h})= ( \vert h_1 \vert \, + \, \vert h_2 \vert ) / \sqrt{2}$.
For any $u \in {\mathcal M}({\bf h})$, i.e., $m({\bf h})  \leq \vert u \vert \leq M({\bf h})$,
the corresponding phase angle vector $\Theta^u = (\theta_1^u \,,\, \theta_2^u)^T$ which satisfies (\ref{u_sat}) is given by
\begin{eqnarray*}
\theta_2^u  =  \cos^{-1}{\Bigg (} \frac{\vert u \vert^2  \, + \, \frac{\vert h_2 \vert^2}{2} \,  - \, \frac{\vert h_1 \vert^2}{2}}  {\sqrt{2} \vert u \vert \vert h_2 \vert} {\Bigg )}  \,\, + \,\, \arg(u) \,\, - \,\, \arg(h_2)  \,\,\,,\,\,\,
\theta_1^u  =  \arg{\Bigg (} \frac{\sqrt{2}}{h_1} {\Big (} u \, - \, \frac{h_2}{\sqrt{2}} e^{j \theta_2^u} {\Big )} {\Bigg )}
\end{eqnarray*}
Note that there can be two possible solutions, since $\cos^{-1}(\cdot)$ can take two possible values in $[-\pi \,\, \pi)$.
}

{
For $N=3$, $M({\bf h}) = ( \vert h_1 \vert \, + \, \vert h_2 \vert \, + \, \vert h_3 \vert) / \sqrt{3}$, and $m({\bf h})$ is given by
{
\vspace{-3mm}
\begin{eqnarray*}
m({\bf h}) = 
\left \{
\begin{array}{cc}
\frac{ {\big \vert} \vert h_1 \vert \, - \, \vert h_2 \vert  {\big \vert}  \, - \, \vert h_3 \vert    }{\sqrt{3}} &  \,\,\, \vert h_3 \vert \,  \leq  \, {\big \vert} \vert h_1 \vert \, - \, \vert h_2 \vert  {\big \vert} \\
0 & \,\,\,  {\big \vert} \vert h_1 \vert \, - \, \vert h_2 \vert  {\big \vert} \,  \leq \, \vert h_3 \vert \, \leq \,  \vert h_1 \vert \, + \, \vert h_2 \vert \\
\frac{ \vert h_3 \vert \, - \, (\vert h_1 \vert \, + \, \vert h_2 \vert) }{\sqrt{3}} & \,\,\, \vert h_3 \vert \, \geq \,   \vert h_1 \vert \, + \, \vert h_2 \vert
\end{array}
\right.
\end{eqnarray*}
}
For any $u \in {\mathcal M}({\bf h})$, i.e., $m({\bf h})  \leq \vert u \vert \leq M({\bf h})$,
the corresponding phase angle vector is $\Theta^u = (\theta_1^u \,,\, \theta_2^u \,,\, \theta_3^u)^T$, with $\theta_3^u$ satisfying
\begin{eqnarray*}
\frac{ 3\vert u \vert^2  + \vert h_3 \vert^2  - (\vert h_1 \vert + \vert h_2 \vert)^2 } {2\sqrt{3} \vert u \vert \vert h_3 \vert}    \leq \cos{\big (}\theta_3^u + \arg(h_3)  - \arg(u){\big )} \leq \frac{ 3\vert u \vert^2  + \vert h_3 \vert^2  - (\vert h_1 \vert - \vert h_2 \vert)^2  } {2\sqrt{3} \vert u \vert \vert h_3 \vert} 
\end{eqnarray*}
Note that, $\theta_3^u$ can take infinitely many values.
For a chosen $\theta_3^u$, let $u_1  \Define  \sqrt{\frac{3}{2}} {\Big (} u \, - \, \frac{h_3 e^{j \theta_3^u}}{\sqrt{3}} {\Big )}$.
The remaining angles are then given by
\begin{eqnarray*}
\theta_2^u  =  \cos^{-1}{\Bigg (} \frac{\vert u_1 \vert^2  \, + \, \frac{\vert h_2 \vert^2}{2} \,  - \, \frac{\vert h_1 \vert^2}{2}}  {\sqrt{2} \vert u_1 \vert \vert h_2 \vert} {\Bigg )}  \,\, + \,\, \arg(u_1) \,\, - \,\, \arg(h_2)   \,\,\,,\,\,\,
\theta_1^u  =  \arg{\Bigg (} \frac{\sqrt{2}}{h_1} {\Big (} u_1 \, - \, \frac{h_2}{\sqrt{2}} e^{j \theta_2^u} {\Big )} {\Bigg )}
\end{eqnarray*}
}
}}
{
From the expressions for the phase angle vector for very small $N$, and the existence of low-complexity gradient-descent type methods
for large $N$, it is expected that the computational complexity of the proposed CE scheme would not be significantly larger than the complexity of
the MRT scheme when $N$ is either very small or large. 
}

When $N$ is neither very small nor large (typically $3 < N \leq 10$), then the value of the error norm function  may not be small at a significant fraction of local minima, which leads
to poor performance of the gradient descent method.
We therefore propose the following two-step algorithm for small $N$ (i.e., $3 < N \leq 10$).
In the first step, we find a value of $\Theta = {\Tilde \Theta^u}$ such that $\vert u - \sum_{i=1}^N h_i e^{j {\Tilde \theta_i^u}} /\sqrt{N} \vert^2$
is {\em sufficiently} small. This step ensures that with high probability, ${\Tilde \Theta^u}$ is
  inside the region of attraction  of the global minimum of the error norm function.
In the second step, with this $\Theta = {\Tilde \Theta^u} =({\Tilde \theta_1^u}, \cdots, {\Tilde \theta_N^u})$ as the initial vector, a simple gradient descent algorithm
would then converge to the global minimum.

The first step of the proposed algorithm is based on the Depth-First-Search (DFS) technique.
Basically, for a given $u$, we start with enumerating the possible values
taken by ${\Tilde \theta_N^u}$ such that (\ref{u_sat}) is satisfied with $\Theta^u = {\Tilde \Theta^u}$.
To satisfy (\ref{u_sat}), it is clear that ${\Tilde \theta_N^u}$ must equivalently satisfy
\begin{equation}
\label{tilde_ueqn}
u - \frac{h_N e^{j {\Tilde \theta_N^u}}} {\sqrt{N}} = \sqrt{\frac{N -1 }{N }}  \frac { \sum_{i=1}^{N - 1}  h_i  e^{j {\Tilde \theta_i^u}} } {\sqrt{N -1}}.
\end{equation}
Using Theorem~\ref{set_Mh}, this is then equivalent to $(\sqrt{N}/\sqrt{N - 1}) (u - \frac {h_N e^{j {\Tilde \theta_N^u}}} {\sqrt{N}}) \in {\mathcal M}((h_1, \cdots, h_{N-1})^T)$
i.e.
\begin{equation}
\label{solve_N}
m({\bf h}^{(N - 1)})  \leq \sqrt{\frac{N }{N - 1}} {\Big \vert} u - \frac {h_N e^{j {\Tilde \theta_N^u}}} {\sqrt{N}} {\Big \vert} \leq M({\bf h}^{(N - 1)})
\end{equation}
where ${\bf h}^{(N - 1)} \Define (h_1, \ldots, h_{N-1})^T$ and $m(\cdot),M(\cdot)$ are defined in (\ref{max_min_def}).
For example $M({\bf h}^{(N - 1)}) = \Vert {\bf h}^{(N - 1)} \Vert_1 / \sqrt{N - 1}$.
Equation (\ref{solve_N}) gives us an {\em admissible} set $I_N^u \subset [-\pi, \pi)$ to which ${\Tilde \theta_N^u}$ must belong for
(\ref{tilde_ueqn}) to be satisfied. We call this as the $k=0$-th ``depth'' level of the proposed DFS technique.

Next, for a given value of ${\Tilde \theta_N^u} \in I_N^u$, we go to the next ``depth'' level (i.e., $k=1$) and find the set
of admissible values for ${\Tilde \theta_{N-1}^u}$. Essentially, at the $k$-th depth level, for a given choice
of values of $({\Tilde \theta_N^u}, {\Tilde \theta_{N-1}^u}, \ldots,  {\Tilde \theta_{N-k+1}^u})$, with ${\Tilde \theta_{N-i+1}^u} \in I_{N -i +1}^u, i=1,\cdots,k$,
we solve for the set of admissible values for ${\Tilde \theta_{N-k}^u}$ such that (\ref{u_sat}) is satisfied with $\Theta^u = {\Tilde \Theta^u}$.
From Theorem \ref{set_Mh}, this set (i.e., $I_{N-k}^u$ ) is given by the values of ${\Tilde \theta_{N-k}^u}$ satisfying
{\small
\begin{eqnarray}
\label{solve_Nmk}
\sqrt{\frac{N - k - 1 }{N }} \,\, m({\bf h}^{(N - k -1)}) \,\,  \leq \,\, {\Big \vert} u^{(k)} - \frac { h_{_{N-k}} e^{j {\Tilde \theta_{N-k}^u}}} {\sqrt{N}}  {\Big \vert} \,\, \leq \,\, \sqrt{\frac{N - k - 1 }{N }} \,\, M({\bf h}^{(N - k - 1)})
\end{eqnarray}
}
\normalsize
where $u^{(k)} \Define (u - \sum_{i=1}^{k} \frac {h_{_{N-i+1}}} {\sqrt{N}}  e^{j {\Tilde \theta_{N-i+ 1}^u}})$ and ${\bf h}^{(N -k- 1)} \Define (h_1, \ldots, h_{N-k-1})^T$.
If there exists no solution to (\ref{solve_Nmk}) (i.e., $I_{N -k}^u$ is empty), then the algorithm backtracks to the previous depth level i.e., $k-1$,
and picks the next possible unexplored admissible value for ${\Tilde \theta_{N-k+1}^u}$ from the set $I_{N -k +1}^u$.
If there exists a solution to (\ref{solve_Nmk}), then the algorithm simply moves to the next depth level, i.e., $k+1$.
The algorithm terminates once it reaches a depth level of $k=N-1$ with a non-empty admissible set $I_1$.
Since $u \in {\mathcal M}({\bf h})$, the algorithm is guaranteed to terminate (by Theorem \ref{set_Mh}).
It can be shown that for depth levels less than $k = N -2$, the admissible set is generally an infinite set (usually a union of
intervals in ${\mathbb R}$). Therefore,
due to complexity reasons, at each depth level it is   suggested to consider only a finite subset
of values from the admissible set (e.g. values on a very fine grid),
and terminate once the algorithm reaches a sufficiently high pre-defined depth level $K$ with
the current error norm $\vert u^{(K)} \vert$ below a pre-defined threshold.
In the second step, a gradient descent algorithm starting with the initial
vector $\Theta = ({\Tilde  \theta_N^u}, \ldots, {\Tilde \theta_{N-K+ 1}^u}, 0 , \ldots, 0)$,
converges to the global minimum of the error norm function $e^u(\Theta)$.
{In terms of complexity, the first step of this two-step algorithm is expected to have a
higher complexity when compared to the MRT scheme.}
\section{The Doughnut Channel}
\begin{figure}[t]
\begin{center}
\hspace{-4mm}
\epsfig{file=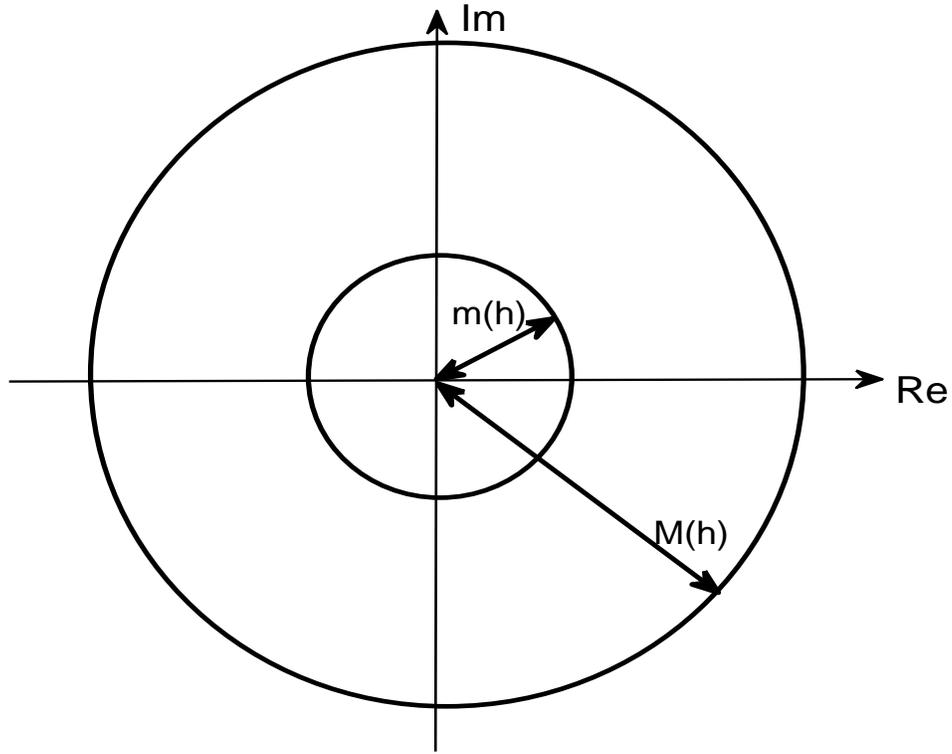, width=130mm,height=100mm}
\end{center}
\vspace{-4mm}
\caption{The doughnut set ${\mathcal M}({\bf h})$ in the complex plane. ${\mathcal M}({\bf h})$ contains all points
in the ``doughnut'' shaped region between the outer and the inner circles of radius $M({\bf h})$ and $m({\bf h})$ respectively.}
\label{fig_donut}
\vspace{-4mm}
\end{figure}

Geometrically the set ${\mathcal M}({\bf h})$ resembles
a ``doughnut'' in the complex plane (see Theorem \ref{set_Mh} and Fig.~\ref{fig_donut}).
With ${\mathcal U} \subseteq {\mathcal M}({\bf h})$,
and the precoder map in (\ref{prec_def}),
we effectively have a ``doughnut channel'' (see (\ref{doughnut_eqn}))
{
\vspace{-3mm}
\begin{equation}
\label{donut_ch}
y = \sqrt{P_T} \,  u + w \,\,\,\,\,\,\,\,,\,\,\,\,\,\,\,\, m({\bf h}) \leq \vert u \vert \leq M({\bf h}) \,\,\,,\,\,\, w \sim {\mathcal C}{\mathcal N}(0,\sigma^2)
\end{equation}
}
which is a SISO AWGN channel where the information symbol $u$ is constrained to belong to the ``doughnut'' set ${\mathcal M}({\bf h})$.
Therefore, with ${\mathcal U} \subseteq {\mathcal M}({\bf h})$, it is clear that the capacity
of the MISO channel with per-antenna CE inputs is equal to the capacity of the doughnut channel in (\ref{donut_ch}),
which is given by
\begin{align}
C_{\footnotesize \mbox{donut}}  =  {\mathop {\sup}  \limits_{p_u(\cdot) \,,\, u \in {\mathcal M}({\bf h})}}   I(y ; u)
\end{align}
where $I(y ; u)$ denotes the mutual information between $y$ and $u$, and $p_u(\cdot)$ is the p.d.f.\
of $u$.
Due to the difficulty in deriving an exact expression for $C_{\footnotesize \mbox{donut}}$,
we propose an appropriate lower and upper bound.
{
The upper and lower bounds presented here will be used in Section \ref{inf_rate_comp_sec} to quantify the performance 
of the proposed CE scheme when compared to the average-only total power constrained MRT scheme.
}
\subsection{An Achievable Information Rate for the Doughnut Channel (Lower Bound on Capacity)}
\label{donut_ch_low_bnd_sec}
For $N=1$, the doughnut set contracts to a circle in the complex plane. In this case, 
capacity is achieved when the input $u$ is uniformly distributed on this circle  \cite{Wyner}.

For $N > 1$, the information rate achieved with $u$ {\em uniformly} distributed inside the doughnut set
(i.e., the p.d.f. of $u$ is $p_u^{\mbox{\footnotesize{unif}}}(z) = \frac {1} {\pi (M({\bf h})^2 - m({\bf h})^2)} \,\,,\,\, z \in {\mathcal M}({\bf h})$) is given by
{
\normalsize
\begin{eqnarray}
\label{mut_inf_bnd}
I(y ; u)^{\mbox{\footnotesize{unif}}} & \hspace{-3mm} = & I{\Big (}\frac{y}{\sqrt{P_T}} ; u{\Big )} =  h{\Big (}\frac{y}{\sqrt{P_T}}{\Big )} - h{\Big (}\frac{y}{\sqrt{P_T}}\,\,  \vert \,\,  u{\Big )} 
=  h{\Big (} u + \frac{w}{\sqrt{P_T}}  {\Big )} - h{\Big (} \frac{w}{\sqrt{P_T}} {\Big )}  \nonumber \\
& \geq & \log_2(2^{h(u)} + 2^{h(w/\sqrt{P_T})}) - h(w/\sqrt{P_T}) 
 =  \log_2(1 + 2^{h(u) - h(w/\sqrt{P_T})})
\end{eqnarray}
}
\normalsize
where $h(s) \Define - \int p_s(z) \log_2(p_s(z)) dz$
denotes the differential entropy of the r.v. $s$ ($p_s(\cdot)$ denotes the p.d.f. of $s$).
The inequality in (\ref{mut_inf_bnd}) follows from the Entropy Power
Inequality (EPI) \cite{Verdu_EPI},
which states that if  $y = u + v$
where   $u$ and $v$ are independent random variables, it holds that
$2^{h(y)} \geq 2^{h(u)} + 2^{h(v)}$.
Since $u$ is uniformly distributed inside ${\mathcal M}({\bf h})$, we have $h(u) = \log_2( \pi (M({\bf h})^2 - m({\bf h})^2))$.
Using this in (\ref{mut_inf_bnd}), {we have the following lower bound}\footnote{\footnotesize{
With $N > 1$, a condition that is required for the usage of EPI to be valid is that $M({\bf h}) > m({\bf h})$, since
otherwise the set
${\mathcal M}({\bf h})$ has a zero Lebesgue measure leading to an undefined $h(u)$.
From Lemma~\ref{max_val} and \ref{min_val_bnd}
it follows that the condition $\Vert {\bf h} \Vert_{1} > \Vert {\bf h} \Vert_{\infty}$ implies
$M({\bf h}) > m({\bf h})$.
Since $\Vert {\bf h} \Vert_{1} > \Vert {\bf h} \Vert_{\infty}$ holds for any ${\bf h}$ having
more than one non-zero component, the required condition is met for most channel fading scenarios of practical interest.}}
\begin{subequations}
\begin{equation}
\label{unif_rate1}
C_{\footnotesize \mbox{donut}} \, \geq \, I(y ; u)^{\mbox{\footnotesize{unif}}}  \geq  \log_2 {\Big (} 1 + \frac{P_T}{\sigma^2} \frac{M({\bf h})^2 -  m({\bf h})^2}{e} {\Big )}
\end{equation}
\begin{equation}
\label{unif_rate2}
C_{\footnotesize \mbox{donut}} \, \geq \, I(y ; u)^{\mbox{\footnotesize{unif}}} \geq  \log_2 {\Big (} 1 + \frac{P_T}{\sigma^2}   \frac{ \Vert {\bf h} \Vert_1^2 - \Vert {\bf h} \Vert_{\infty}^2} {N e} {\Big )}
\,\,\,\,\mbox{\footnotesize{(using Lemmas \ref{max_val},\ref{min_val_bnd})}}.
\end{equation}
\end{subequations}
\subsection{An Upper Bound on the Doughnut Channel Capacity}
\label{donut_ch_upp_bnd_sec}
Let $s \Define \frac{y}{\sqrt{P_T}} $, and let $p_s(\cdot)$ be its p.d.f.
We now have
\begin{eqnarray}
\label{upp_bnd_eqn}
I(y ; u) & = & I(s ; u) = h(s) - h(s \,| \, u) 
 =  - \int_{z \in {\mathbb C}} p_s(z) \log_2(p_s(z)) dz \, - \, \log_2{\Big (}\pi e \frac{\sigma^2}{P_T}{\Big )} \nonumber \\ 
&  = & - \int_{z \in {\mathbb C}}  p_s(z) \log_2{\Big (}\frac{p_s(z)}{g(z)}{\Big )} dz -  \int_{z \in {\mathbb C}}  p_s(z) \log_2({g(z)}) dz  \, - \, \log_2{\Big (}\pi e \frac{\sigma^2}{P_T}{\Big )} \nonumber \\
& = & -{\mbox D}(p_s(.) \vert \vert g(.)) - \int_{z \in {\mathbb C}}  p_s(z) \log_2({g(z)}) dz  \, - \, \log_2 {\Big (}\pi e \frac{\sigma^2}{P_T}{\Big )} \nonumber \\
& \leq & - \int_{z \in {\mathbb C}}  p_s(z) \log_2({g(z)}) dz  \, - \, \log_2{\Big (}\pi e \frac{\sigma^2}{P_T}{\Big )}
\end{eqnarray}
where $g(z)$ is some distribution function
(i.e., $\int_{z \in {\mathbb C}} g(z) dz \,=\, 1$).
Also, for any $z \in {\mathbb C}$, $g(z) > 0$.
${\mbox D}(p_s(\cdot) \vert \vert g(\cdot))$
denotes the Kullback-Leibler (KL) distance
between the distributions $p_s(\cdot)$ and $g(\cdot)$.
The last inequality in (\ref{upp_bnd_eqn}) follows from the fact that
the KL distance between any two distributions is always non-negative. Since (\ref{upp_bnd_eqn}) holds for any
distribution $g(\cdot)$, we aim to find a $g(\cdot)$ for which the integral $\int_{z \in {\mathbb C}}  p_s(z) \log_2({g(z)}) dz $
can be computed in closed-form, and which also results in a sufficiently tight upper bound.
We propose to use $g(z) = 2 \beta e^{- \pi^3 \beta^2 \vert z \vert^4}$, $\beta > 0$. With this choice of $g(z)$ in (\ref{upp_bnd_eqn}), we have
\begin{eqnarray}
\label{Iuy_finalbetabnd}
I(y ; u) & \leq & -\log_2(2 \beta) + \pi^3 \beta^2 \log_2(e)
{\Big (} M({\bf h})^4 +  2 \frac{\sigma^4}{P_T^2} + 4 \frac{\sigma^2}{P_T} M({\bf h})^2
   {\Big )} - \log_2(\pi e \frac{\sigma^2}{P_T}).
\end{eqnarray}
Minimizing this upper bound w.r.t. the free parameter $\beta > 0$ gives
\begin{eqnarray}
\label{Iuy_finalbetabnd2}
I(y ; u) & \leq & I^{(1)} {\Big (} {\bf h},\frac{P_T}{\sigma^2} {\Big )} \,\,,\,\,
I^{(1)} {\Big (}{\bf h},\frac{P_T}{\sigma^2} {\Big )}   \Define   \frac{1}{2} \log_2 {\Big (} \frac{\pi}{2 e}{\Big )} + \frac{1}{2} \log_2{\Big (} M({\bf h})^4 {\Big (}\frac{P_T}{\sigma^2}{\Big )}^2 + 4 M({\bf h})^2 {\Big (}\frac{P_T}{\sigma^2}{\Big )} +  2   {\Big )} \nonumber \\
& \leq & \frac{1}{2} \log_2 {\Big (} \frac{2 \pi}{e} {\Big )} + \log_2{\Big (} 1 + \frac{P_T}{\sigma^2} \frac{M({\bf h})^2}{2}  {\Big )}
\end{eqnarray}
The bound in (\ref{Iuy_finalbetabnd2}) is always valid irrespective of the distribution
of $u$. Therefore it holds also for the distribution of $u$ which maximizes $I(y ; u)$ subject to $u \in {\mathcal M}({\bf h})$.

Another upper bound to $C_{\footnotesize \mbox{donut}}$ is given by
the capacity of a MISO channel where the per-antenna {\em average-only} power is constrained to be $P_T/N$
(i.e., ${\mathbb E}[ \vert x_i \vert^2] = P_T/N\,\,,i=1,\ldots,N$) for every channel realization ${\bf h}$.
We shall subsequently refer to this constraint as PAPC.
The capacity of the MISO channel
under a PAPC constraint is given by \cite{Mai}\footnote{\footnotesize{
For the PAPC constrained MISO channel, capacity is achieved by choosing $u$ to be Gaussian distributed unit-energy symbols. For a given symbol $u$ to be communicated, the optimal PAPC precoder transmits $\sqrt{P_T/N} ({h_i^*}/|h_i|) u$ from the $i$-th antenna.}}
{
\vspace{-6mm}
\begin{eqnarray}
\label{PAPC_cap}
C_{\footnotesize \mbox{PAPC}} & = & \log_2 {\Big (} 1 + \frac{P_T}{\sigma^2} M({\bf h})^2 {\Big )}
\end{eqnarray}
}
It is clear that, for a given total transmit power $P_T$,
the PAPC constraint is much less restrictive than the CE constraint,
and therefore $C_{\footnotesize \mbox {donut}} \leq C_{\footnotesize \mbox{PAPC}}$.
We finally propose the following upper bound  
on $C_{\footnotesize \mbox{donut}}$
{
\vspace{-2mm}
\begin{eqnarray}
\label{Iuy_finalbetabnd3}
C_{\footnotesize \mbox{donut}} & \leq  &  I^{(2)} {\Big (} {\bf h},\frac{P_T}{\sigma^2} {\Big )}  \,\,\,,\,\,\,
I^{(2)} {\Big (} {\bf h},\frac{P_T}{\sigma^2} {\Big )}  \Define     \min {\Big (} I^{(1)} {\Big (} {\bf h},\frac{P_T}{\sigma^2} {\Big )}  \,,\, C_{\footnotesize \mbox{PAPC}}     {\Big )}
\end{eqnarray}
}
where $I^{(1)} {\Big (} {\bf h},\frac{P_T}{\sigma^2} {\Big )} $ has been defined in (\ref{Iuy_finalbetabnd2}).
\section{On the capacity achieving input distribution for the doughnut channel}
\label{cap_ach_inp_dist}
For the i.i.d. Rayleigh fading channel, i.i.d. fading channels with bounded channel gain and the DLOS channel, with high probability, the inner radius of the doughnut set ${\mathcal M}({\bf h})$ shrinks to $0$ as
$N \rightarrow \infty$ (see Section \ref{set_study}). 
This implies that, for large $N$ the doughnut channel in (\ref{donut_ch}) is essentially a 
{\em peak-input-amplitude only} limited SISO AWGN channel, with the per-channel use peak-amplitude constraint
$\vert u \vert \leq M({\bf h})$, i.e.
\begin{eqnarray}
y & = & \sqrt{P_T} u + w \,\,\,\,\,\,,\,\,\,\,\,\,  \vert u \vert \leq {M}({\bf h}) \,\,\,,\,\,\, w \sim {\mathcal C}{\mathcal N}(0,\sigma^2) .
\end{eqnarray}
In the following, for large $N$ we exploit this observation to propose a near-optimal capacity achieving input distribution ($p_u(\cdot)$) for the doughnut channel.

In \cite{Shamai}, it has been shown that, for a peak-input-amplitude only constrained SISO AWGN channel,
capacity is achieved
with channel inputs that are {\em discrete} in {\em amplitude} and {\em uniform in phase} (DAUIP).
In our notation, the information symbol $u \in {\mathcal U}_{\footnotesize \mbox{DAUIP}}^{L,\alpha}$, where
${\mathcal U}_{\footnotesize \mbox{DAUIP}}^{L,\alpha} = \cup_{l=1}^{L} {\mathcal U}_{\footnotesize \mbox{DAUIP}}^l$,
with $L \in {\mathbb Z}^+$ and $\alpha = (\alpha_1,\alpha_2,\cdots,\alpha_L)^T$
($\alpha_l  \in (0 \,,\, 1] \,,\,$$\alpha_1 < \alpha_2 < \cdots < \alpha_L \leq 1$).
${\mathcal U}_{\footnotesize \mbox{DAUIP}}^l$ is given by
{
\vspace{-3mm}
\begin{eqnarray}
{\mathcal U}_{\footnotesize \mbox{DAUIP}}^l = \{ v \in {\mathbb C} \,\,| \,\, \vert v \vert = \alpha_l { M}({\bf h}) \}.
\vspace{-2mm}
\end{eqnarray}
}
Essentially, the DAUIP alphabet set is composed of $L$ circles in ${\mathbb C}$ with the $l$-th circle
having amplitude $\alpha_l {M}({\bf h}), l=1,2,\ldots,L$.
Furthermore, within a given circle, each point is equally likely (i.e., the phase is uniformly distributed).
Let the probability that the information symbol $u$ belongs to the $l$-th circle be denoted by
$p_l\,,\,l=1,2,\ldots,L$, $\sum_l p_l = 1$.
In \cite{Shamai}, no closed-form expressions were given, neither for the capacity nor for
the capacity achieving input (i.e., $L$, $\{ \alpha_l \}$ and $\{ p_l \}$).
However, in \cite{Shamai}, numerically it was shown that, at low peak-SNR
(i.e., low $(P_T/\sigma^2) {M}({\bf h})^2$ in our notation), it is optimal to use a single-amplitude DAUIP alphabet set with $L=1 \, , \,\alpha_1 = 1$,
whereas with increasing peak-SNR,
the number of circles in the optimal DAUIP alphabet also increases.

Based on the above discussion, for i.i.d. Rayleigh fading channel, i.i.d. fading channel with bounded channel gains and DLOS channels it can be concluded that,  at large $N$,
DAUIP inputs/alphabets are nearly optimal in terms of achieving the capacity of the doughnut channel/CE constrained MISO channel.
In this paper, for a given $N$ and $P_T/\sigma^2$, we numerically optimize the ergodic mutual information of the doughnut channel
w.r.t.\ $L$ and $\alpha_1 < \alpha_2 < \cdots < \alpha_L \leq 1$, i.e.
\begin{eqnarray}
\label{l_star}
(L^{\star},\alpha^\star) & \Define & \arg \max_{L \in {\mathbb Z}^+, 0 < \alpha_1 < \cdots < \alpha_L \leq 1} {\mathbb E}_{{\bf h}} [ I(y ; u) ] 
\end{eqnarray}
where, for a given ($L, \alpha$), $u \in {\mathcal U}_{\footnotesize \mbox{DAUIP}}^{L,\alpha}$ and\footnote{\footnotesize{To be precise, ${\mathcal U}_{\footnotesize \mbox{DAUIP}}^l$
is chosen to consist of all complex numbers having magnitude $m({\bf h}) + \alpha_l (M({\bf h}) - m({\bf h}) )$. This choice
is motivated by the fact that, for finite $N > 1$,
even though $m({\bf h})$ is small compared to $M({\bf h})$, it is not exactly $0$.}} $p_1=p_2=\cdots=p_L=1/L$.\footnote{\footnotesize{
In general, $p_1=p_2=\cdots=p_L=1/L$ need not be optimal in terms of maximizing the ergodic mutual information. However, for the i.i.d.\ Rayleigh
fading channel, we numerically observed that, in the practically interesting regime of low to moderate peak-SNR, it was optimal to have only a single
circle, i.e., $L=1$ (for which the trivial probability distribution is $p_1 = 1$). Also, designing practical channel codes for the doughnut channel
would be much simpler when
$p_1=p_2=\cdots=p_L=1/L$.}}
The numerical optimization in (\ref{l_star}) can be performed off-line and therefore does not impact the online precoding complexity.

\section{Information rate comparison}
\label{inf_rate_comp_sec}
With an average-only total transmit power constraint (ATPC), MRT with Gaussian information alphabet achieves the capacity of the
single user Gaussian MISO channel, which is given by
{
\vspace{-1mm}
\begin{equation}
\label{C_MRT}
C_{\footnotesize \mbox{ATPC}} = \log_2 {\Big (} 1 + \Vert  {\bf h} \Vert_{_2}^2 \frac{P_T}{\sigma^2} {\Big )}.
\end{equation}
}
Comparing (\ref{C_MRT}) with (\ref{PAPC_cap}) and (\ref{Iuy_finalbetabnd3}) we have
{
\vspace{-3mm}
\begin{eqnarray}
\label{cap_cmp_eq}
C_{\footnotesize \mbox{donut}} < C_{\footnotesize {\mbox PAPC}} \leq  C_{\footnotesize \mbox{ATPC}}
\end{eqnarray}
}
For a desired information rate $R$, let the ratio of the total transmit power required under the per-antenna CE constraint
to the total transmit power required under ATPC be referred to as the ``power gap''
between the proposed CE precoder and the MRT precoder (denoted by $P_{\footnotesize \mbox{gap}}^{\footnotesize \mbox{CE},\mbox{MRT}}(R)$
).\footnote{\footnotesize{
In the following, we drop the argument $R$ for notational brevity.}}
{
\begin{eqnarray}
P_{\footnotesize \mbox{gap}}^{\footnotesize \mbox{CE},\mbox{MRT}}(R) & \Define & \frac{P^{\footnotesize \mbox{CE}}(R)}{P^{\footnotesize \mbox{MRT}}(R)}
\end{eqnarray}
where $P^{\footnotesize \mbox{CE}}(R)$ and $P^{\footnotesize \mbox{MRT}}(R)$ denote the total transmit power required by the CE scheme and the MRT scheme respectively, to achieve information rate $R$.
}
We can similarly define the power gap between the proposed CE precoder and an optimal precoder operating under the PAPC constraint (denoted by
$P_{\footnotesize \mbox{gap}}^{\footnotesize \mbox{CE},\mbox{PAPC}}$).
In the following, we investigate the power gap and the capacity ratios between the proposed CE precoder and the PAPC, MRT precoders, at low and high $P_T/\sigma^2$ (results are summarized in Table~\ref{table_0}).
\vspace{-2mm}
\subsection{Information Rate Comparison at Low $P_T M({\bf h})^2 /\sigma^2$}
\label{sec_low_snr}
From the discussion in Section \ref{cap_ach_inp_dist} we know that, at low $P_T M({\bf h})^2 /\sigma^2$ and large $N$, 
a single amplitude DAUIP  information alphabet having complex symbols of magnitude $M({\bf h})$ achieves near-capacity performance for the doughnut channel.
Therefore, {\em in the low $P_T M({\bf h})^2 /\sigma^2$ regime, the capacity of the doughnut channel is roughly equal to that of a SISO non-fading AWGN channel (noise variance $\sigma^2$),
where the input is constrained to have a constant envelope/amplitude of $M({\bf h}) \sqrt{P_T}$}, i.e.,
\begin{eqnarray}
\label{low_snr_donut_ch}
y & = &  u + w \,\,,\,\, \vert u \vert = M({\bf h}) \sqrt{P_T} \,,\, w \sim {\mathcal C}{\mathcal N}(0,\sigma^2).
\end{eqnarray}
The CE input constrained SISO AWGN channel in (\ref{low_snr_donut_ch}) was considered by Wyner in \cite{Wyner}.
In \cite{Wyner}, it was shown that for an average power only constrained AWGN channel (i.e., $y = u + w$),
{\em using a CE input (instead of the capacity optimal
Gaussian input) is almost information lossless for
$\mbox{SNR} = {\mathbb E}[\vert u \vert^2]/\sigma^2 \leq 1$ }.\footnote{\footnotesize
See Eq.~(14) and Fig.~$2$ in \cite{Wyner}.} Hence for $P_T M({\bf h})^2 /  \sigma^2  \leq 1$ the capacity of the
channel in (\ref{low_snr_donut_ch}) is roughly $\log_2(1 + P_T M({\bf h})^2  / \sigma^2)$.
Using the capacity equivalence between the doughnut channel and the
channel in (\ref{low_snr_donut_ch}), we have
\begin{eqnarray}
\label{donut_cap_low_snr}
C_{\footnotesize \mbox{donut}} \approx \log_2(1 + \frac{P_T}{\sigma^2} M({\bf h})^2) \,\,\,\,\,\,\,\,\,\,\, \mbox{for}\,\,\,  \frac{P_T}{\sigma^2} M({\bf h})^2 \leq 1.
\end{eqnarray}
Further, comparing (\ref{donut_cap_low_snr}) with (\ref{PAPC_cap}), we finally arrive at the conclusion that
at low $ (P_T/\sigma^2) M({\bf h})^2 \leq 1$
\begin{eqnarray}
\label{donut_eq_PAPC}
C_{\footnotesize \mbox{donut}} \approx  C_{\footnotesize \mbox{PAPC}} \,\,\,\,\,\,\,\,\,\,\, \mbox{for} \,\,\, \frac{P_T}{\sigma^2} M({\bf h})^2 \leq 1.
\end{eqnarray}
Note that (\ref{donut_eq_PAPC}) holds for {\em any} channel realization ${\bf h}$.
Using the capacity expressions (\ref{C_MRT}) and (\ref{donut_cap_low_snr}), we can now conclude that,
{
\vspace{-4mm}
\begin{eqnarray}
\label{ce_mrt_pgap}
P_{\footnotesize \mbox{gap}}^{\footnotesize \mbox{CE},\mbox{MRT}} & \approx & \Vert {\bf h} \Vert^2 \, / \,  M({\bf h})^2
\,\, = \,\, \frac {\frac{\sum_{i=1}^N \vert h_i \vert^2}{N} } { {\Big (} \frac{\sum_{i=1}^N \vert h_i \vert }{N} {\Big )}^2} \geq 1   \,\,\,\,\,\,  \mbox{for} \,\,\, \frac{P_T}{\sigma^2} M({\bf h})^2 \leq 1.
\end{eqnarray}
}
{
It is therefore clear that, at low SNR the power gap will be small when the channel gains from each antenna
are similar in magnitude, and the power gap can be large when there is a large variation in the channel gains.
}
For the single-path DLOS channel with
$\vert h_1 \vert = \cdots = \vert h_N \vert$, $P_{\footnotesize \mbox{gap}}^{\footnotesize \mbox{CE},\mbox{MRT}} \approx 1$ for any ${\bf h}$.
For i.i.d. Rayleigh fading channel and i.i.d. fading channels with bounded channel gains, using the law of large numbers and the Slutsky's Theorem, it can be shown that as $N \rightarrow \infty$ 
\begin{eqnarray}
\label{pgap_asymp}
P_{\footnotesize \mbox{gap}}^{\footnotesize \mbox{CE},\mbox{MRT}} \approx \frac {\frac{\sum_{i=1}^N \vert h_i \vert^2}{N} } { {\Big (} \frac{\sum_{i=1}^N \vert h_i \vert }{N} {\Big )}^2}  \rightarrow_p \frac{{\mathbb E}[\vert h_i \vert^2]}{{\big (}{\mathbb E}[\vert h_i \vert]{\big )}^2}\,\,\,\,\,\,\, 
\end{eqnarray}
where $\rightarrow_p$ denotes convergence in probability (w.r.t. the distribution of ${\bf h}$).
For i.i.d. Rayleigh fading, this asymptotic (in $N$) power gap limit is $10 \log_{10}({\mathbb E}[\vert h_i \vert^2]/{\big (}{\mathbb E}[\vert h_i \vert]{\big )}^2) = 1.05$ dB.

\begin{figure}[t]
\begin{center}
\hspace{-4mm}
\epsfig{file=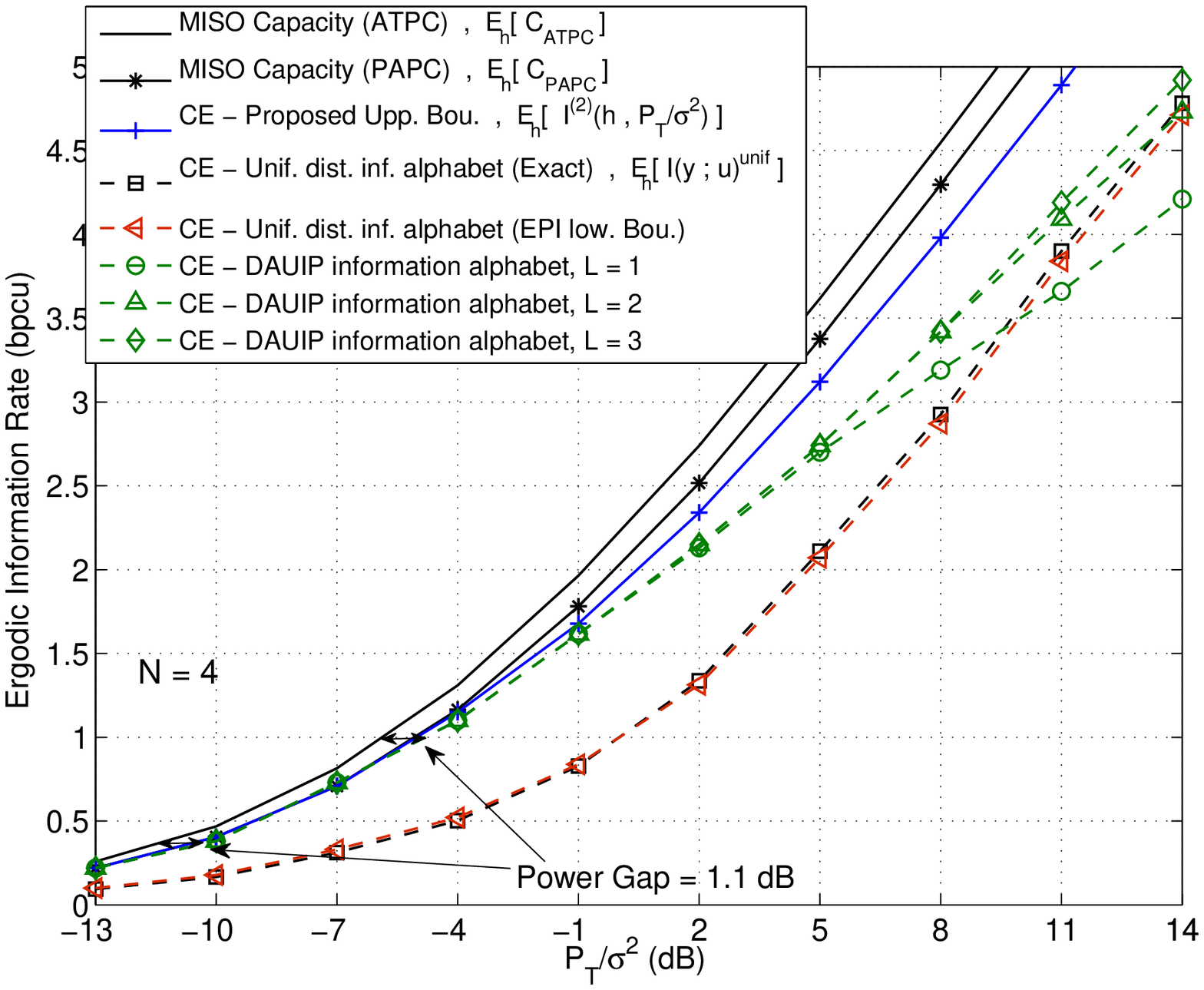, width=155mm,height=113mm}
\end{center}
\vspace{-4mm}
\caption{Ergodic information rate vs.\ $P_T/\sigma^2$, for i.i.d. ${\mathcal C}{\mathcal N}(0,1)$ Rayleigh fading and $N=4$.}
\label{fig_N4_ce_cap}
\vspace{-4mm}
\end{figure}

As an illustrative numerical example, for the i.i.d. Rayleigh fading channel (with ${\mathcal C}{\mathcal N}(0,1)$ distributed channel gains),
in Figs.~ \ref{fig_N4_ce_cap} and \ref{fig_N64_ce_cap}, we plot the ergodic information rate achieved under
the ATPC, PAPC and CE input constraints for $N=4$ and $N = 64$ respectively (as a function of $P_T/\sigma^2$).
In both  figures, for the proposed CE precoder with a DAUIP alphabet,
we plot the ergodic information rate for different fixed values of $L$ (i.e., $L$ is fixed and does not change
with $P_T/\sigma^2$ or with ${\bf h}$). For a fixed $L$ and a given $P_T/\sigma^2$, we numerically maximize the achievable ergodic information rate
as a function of $\alpha=(\alpha_1,\cdots,\alpha_L)$. Note that $\alpha$ only varies with $P_T/\sigma^2$, and does not vary with ${\bf h}$.  For the special case of $L=1$, we always choose $\alpha_1 = 1$.
From the figures, it can be observed that, indeed at low $P_T M({\bf h})^2 /\sigma^2 \leq 1$ (corresponding to achievable rates $\log_2(1 + P_T M({\bf h})^2 /\sigma^2) \leq 1$ bpcu), as discussed previously,
the information rate achieved by the proposed CE precoder with a single amplitude DAUIP information alphabet ($L =1$) equals the MISO capacity under PAPC.
This confirms (\ref{donut_eq_PAPC}), and also shows that the single amplitude DAUIP information alphabet ($L =1$)
is near-optimal for the proposed CE precoder at low $P_T M({\bf h})^2/\sigma^2 \leq 1$.
Note that at low $P_T  M({\bf h})^2 /\sigma^2$, the ergodic information rate achieved with an information alphabet uniformly distributed inside the doughnut set,
is strictly sub-optimal.
Also, at low $P_T M({\bf h})^2 /\sigma^2 \leq 1$, the power gap of the proposed CE precoder (DAUIP, $L = 1$) from the ATPC constrained MRT precoder
is about $1.1$ dB (close to the asymptotic power gap limit of
$1.05$ dB, see (\ref{pgap_asymp})). Note that, even with small $N=4$, the CE-MRT power gap is close to the asymptotic limit.

Note that at low $P_T/\sigma^2$, we have $C_{\footnotesize \mbox{ATPC}} = \log_2(1 + (P_T/\sigma^2) \Vert {\bf h} \Vert^2) \approx (P_T/\sigma^2) \Vert {\bf h} \Vert^2 \log_2(e)$.
Similarly, $C_{\footnotesize \mbox{donut}} \approx (P_T/\sigma^2)  M({\bf h})^2 \log_2(e)$.
Therefore for low $P_T/\sigma^2$, we have
\begin{eqnarray}
\label{c_ratio}
\frac {C_{\footnotesize \mbox{donut}}}{C_{\footnotesize \mbox{ATPC}}} & \approx &  \frac {M({\bf h})^2 }{ \Vert {\bf h} \Vert^2} \,\,\,\,\,\,\,\,\, \mbox{for} \,\,\, \frac{P_T}{\sigma^2} M({\bf h})^2 \ll 1.
\end{eqnarray}
which converges to
$({\mathbb E}[\vert h_i \vert])^2/{\mathbb E}[ \vert h_i \vert^2]$ as $N \rightarrow \infty$ for the i.i.d. Rayleigh fading channel and i.i.d. fading channels with bounded channel gains.
For the single-path DLOS channel this ratio is $1$, i.e., per-antenna CE transmission is optimal
even under ATPC.
For the special case of $N=1$, from \cite{Wyner} it follows that at low $P_T \vert h_1 \vert^2/\sigma^2 \ll 1$,
$C_{\footnotesize \mbox{donut}} \approx  C_{\footnotesize \mbox{ATPC}}$.
\vspace{-2mm}
\subsection{Information Rate Comparison at High $P_T M({\bf h})^2/\sigma^2$}
\label{sec_high_snr}
\begin{figure}[t]
\begin{center}
\hspace{-4mm}
\epsfig{file=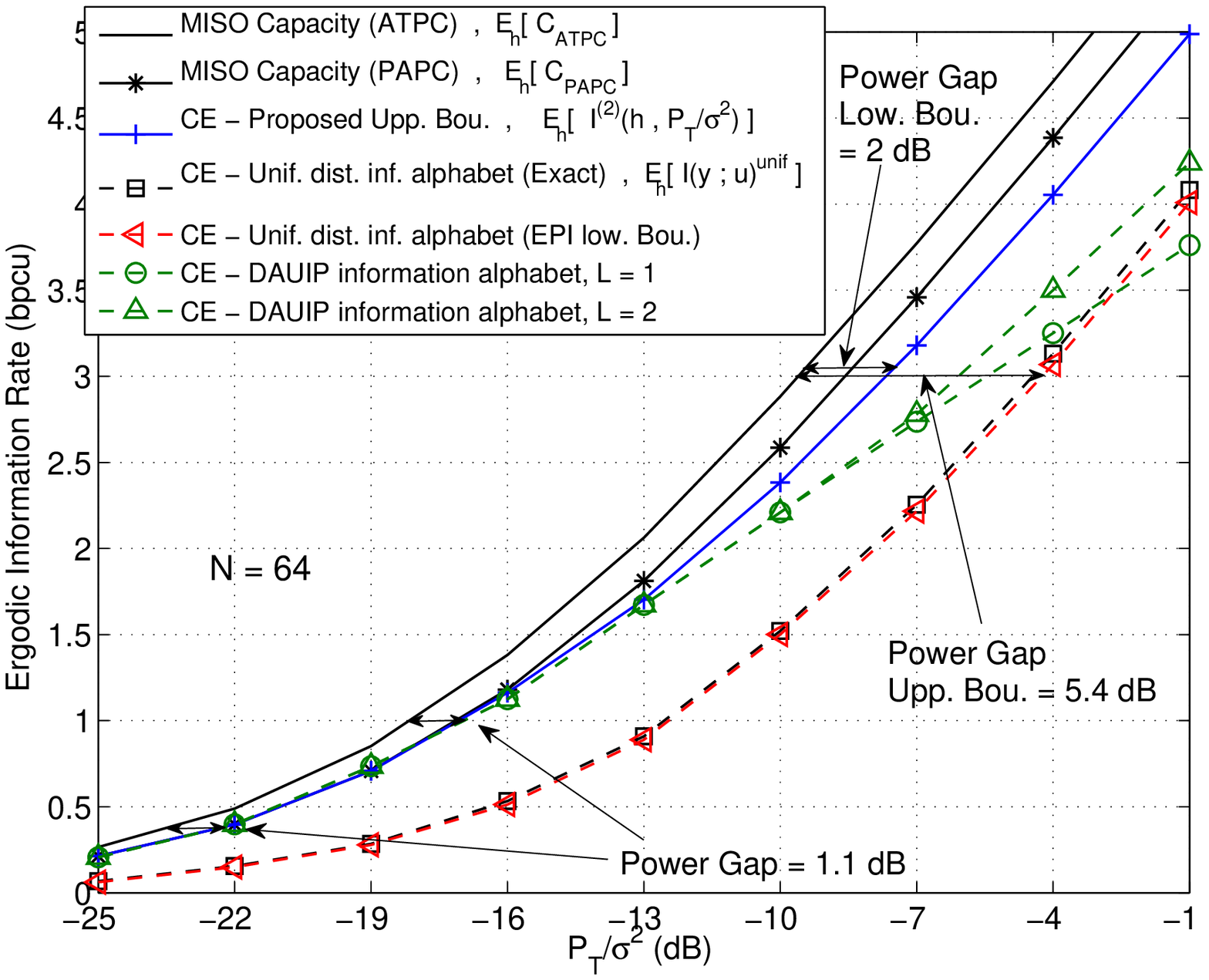, width=160mm,height=107mm}
\end{center}
\vspace{-4mm}
\caption{Ergodic information rate vs. $P_T/\sigma^2$ for i.i.d. ${\mathcal C}{\mathcal N}(0,1)$ Rayleigh fading and $N=64$.}
\label{fig_N64_ce_cap}
\vspace{-4mm}
\end{figure}

\begin{figure}[t]
\begin{center}
\hspace{-4mm}
\epsfig{file=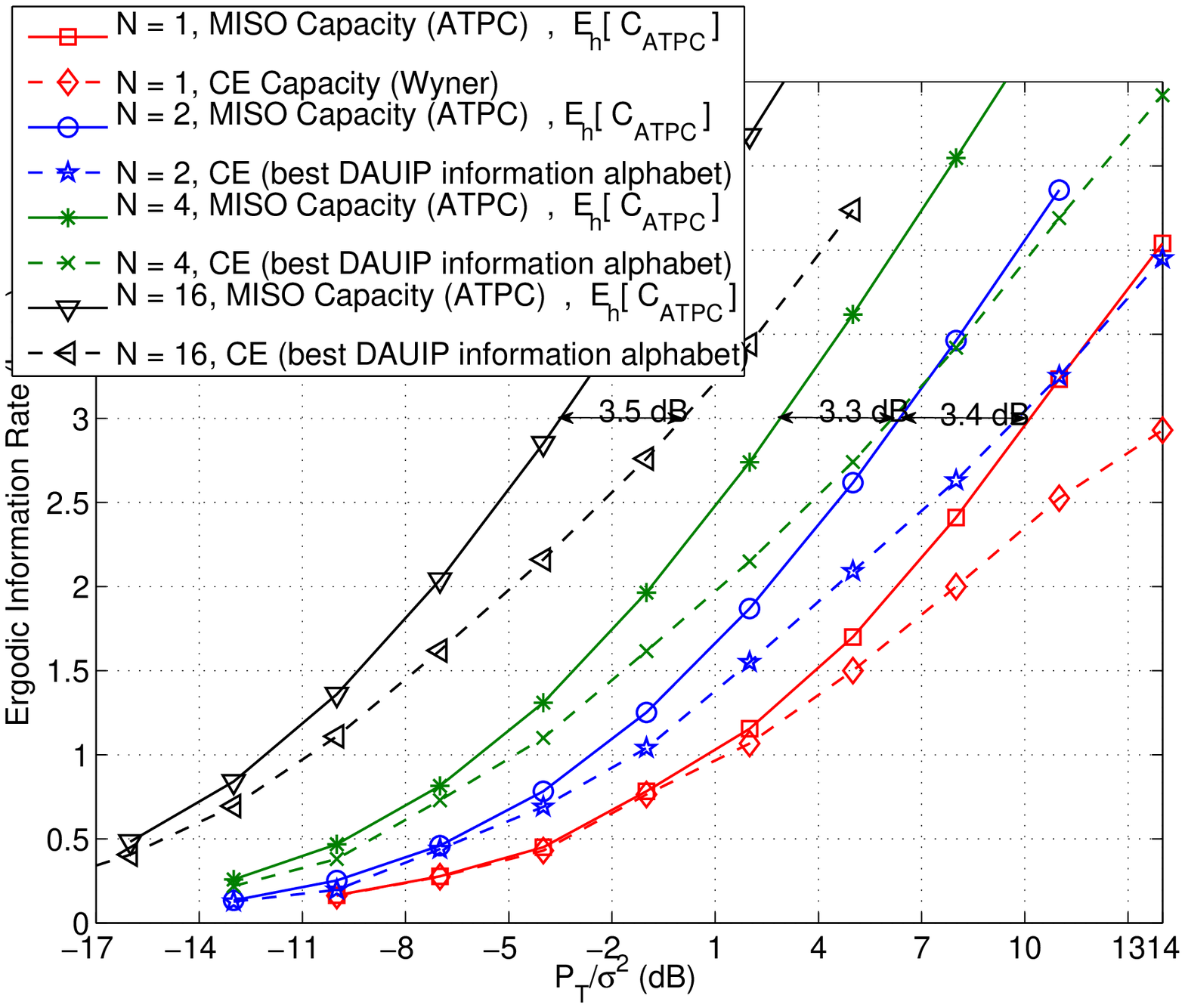, width=155mm,height=110mm}
\end{center}
\vspace{-4mm}
\caption{Ergodic information rate vs. $P_T/\sigma^2$ for i.i.d. ${\mathcal C}{\mathcal N}(0,1)$ Rayleigh fading and $N=1,2,4,16$.}
\label{fig_N116_ce_cap}
\vspace{-4mm}
\end{figure}

In this section we derive lower and upper bounds to the CE-MRT power gap at high $P_T M({\bf h})^2/\sigma^2$.
Using the upper bound to the doughnut channel capacity in (\ref{Iuy_finalbetabnd2}), it follows that in the asymptotic power limit as $P_T M({\bf h})^2 /\sigma^2 \rightarrow \infty$, the CE-MRT power gap
is lower bounded as
\begin{eqnarray}
\label{low_bnd_pgap_ce_mrt}
P_{\footnotesize \mbox{gap}}^{\footnotesize \mbox{CE},\mbox{MRT}} &  \geq & \frac {2 \Vert {\bf h} \Vert^2} { M({\bf h})^2} \,\,\,\,\,\,\, \mbox{for} \,\,\, \frac{ P_T}{\sigma^2} M({\bf h})^2 \gg 1.
\end{eqnarray}
For single-path DLOS channels, this lower bound on the CE-MRT power gap equals $3$ dB,
while for the i.i.d. Rayleigh fading channel and i.i.d. channels with bounded channel gains, it converges to $2 {\mathbb E}[ \vert h_i \vert^2]/({\mathbb E}[\vert h_i \vert])^2$
as $N \rightarrow \infty$ (this is $4.06$ dB for i.i.d. Rayleigh fading channel).
Another interesting fact is that, at high $ \frac{ P_T}{\sigma^2} M({\bf h})^2$, comparing the doughnut channel capacity upper bound in (\ref{Iuy_finalbetabnd2}) and the
PAPC capacity in (\ref{PAPC_cap}) reveals that for {\em any} channel realization ${\bf h}$ and any $N$,
{
\vspace{-4mm}
\begin{eqnarray}
P_{\footnotesize \mbox{gap}}^{\footnotesize \mbox{CE},\mbox{PAPC}} & \geq & 2 \,\,\,\,\,\,\, \mbox{for} \,\,\, \frac{ P_T}{\sigma^2} M({\bf h})^2 \gg 1.
\end{eqnarray}
}
We now obtain an upper bound on the CE-MRT power gap.
Using (\ref{unif_rate1}) and (\ref{C_MRT}) it follows that, for any $P_T  M({\bf h})^2 / \sigma^2$ (not necessarily high),
the CE-MRT power gap can be upper bounded as
{
\begin{eqnarray}
\label{upp_bnd_pgap_ce_mrt}
P_{\footnotesize \mbox{gap}}^{\footnotesize \mbox{CE},\mbox{MRT}}  & \leq & \frac{1}{\kappa} \,\,\,\,\,\,\mbox{\small (for all $ \frac{ P_T}{\sigma^2} M({\bf h})^2$)} \,\,,\,\,
\kappa  \Define  \frac{ M({\bf h})^2 - m({\bf h})^2} { e \Vert {\bf h} \Vert_2^2 }
= \frac { {\Big (} \frac {\sum_i \vert h_i \vert } {N}   {\Big )}^2 - \frac {m({\bf h})^2 } {N}   }   {e \frac {\sum_i \vert h_i \vert^2 } {N}    }
\end{eqnarray}
}
For a single-path only DLOS channel with $\vert h_1 \vert = \ldots = \vert h_N \vert$,
for any ${\bf h}$ and any $N > 1$ it can be shown that $1/\kappa = e$ (since $m({\bf h}) = 0$ for $N > 1$).
With i.i.d. Rayleigh fading and i.i.d. fading channels with bounded channel gains, as $N \rightarrow \infty$,
using the law of large numbers and Slutsky's Theorem  along with the fact that $m({\bf h})\to 0$   as
$N \rightarrow \infty$ (see Section \ref{set_study}),
{
\vspace{-3mm}
\begin{equation}
\label{k_limit}
\kappa \rightarrow_p {({\mathbb E}[ \vert h_i \vert])^2} \, / \, { e {\mathbb E}[ \vert h_i \vert^2]}
\end{equation}
}
Therefore, for the i.i.d. Rayleigh fading channel and i.i.d. fading channels with bounded channel gains, in the asymptotic limit as $N \rightarrow \infty$,
combining (\ref{low_bnd_pgap_ce_mrt}), (\ref{upp_bnd_pgap_ce_mrt}) and (\ref{k_limit}) we have,
\begin{eqnarray}
\label{asymp_upp_low_bnd_ce_mrt_high}
2 \frac {  {\mathbb E}[ \vert h_i \vert^2]} {({\mathbb E}[ \vert h_i \vert])^2} \,\, \leq \,\, P_{\footnotesize \mbox{gap}}^{\footnotesize \mbox{CE},\mbox{MRT}}
\,\, \leq \,\, e \frac {  {\mathbb E}[ \vert h_i \vert^2]} {({\mathbb E}[ \vert h_i \vert])^2} \,\,\,\,\, {\Big (} N \gg 1\,\,,\,\,\frac{P_T}{\sigma^2}  M({\bf h})^2 \gg 1 \,\,{\Big )}
\end{eqnarray}
Therefore, with sufficiently large $N$ and high $P_T M({\bf h})^2 /\sigma^2$,
the difference between the upper and the lower bounds on the CE-MRT power gap
is $10\log_{10}(e/2) = 1.33$ dB {\em irrespective of the channel fading distribution} (as long as the channel gains are bounded).
For i.i.d. Rayleigh fading, the asymptotic upper and lower bounds on the CE-MRT power gap are
$5.4$ and $4.1$ dB respectively, see Fig.~\ref{fig_N64_ce_cap}.\footnote{\footnotesize{
In Fig.~\ref{fig_N64_ce_cap}, note that the power gap lower bound at a desired information rate of $3$ bpcu is only $2$ dB, as compared to the
power gap lower bound limit of $4.1$ dB.
This is because, for a desired rate of $3$ bpcu, the corresponding $P_T M({\bf h})^2/\sigma^2$ is still not high enough for the asymptotic
lower bound in (\ref{asymp_upp_low_bnd_ce_mrt_high}) to be valid. 
A stronger result which can be seen by comparing (\ref{low_bnd_pgap_ce_mrt}) and (\ref{upp_bnd_pgap_ce_mrt}) is that, for channels where $m({\bf h}) \rightarrow 0$ as $N \rightarrow \infty$, at high $P_T M({\bf h})^2/\sigma^2$ the upper to lower bound gap is $1.33$ dB {\em for any ${\bf h}$} (not limited to i.i.d. fading).
}} 
For the practically interesting low to moderate $P_T M({\bf h})^2/\sigma^2$ regime, with DAUIP alphabets the CE-MRT
power gap is usually lesser than its asymptotic lower bound.
We illustrate this fact through Fig.~\ref{fig_N116_ce_cap}, where we plot the ergodic information rate as a function of increasing $P_T/\sigma^2$ for the MRT
and the proposed CE precoder (i.i.d. Rayleigh fading channel). The reported ergodic rate for the proposed CE precoder is with the proposed best DAUIP information alphabet
in (\ref{l_star}).
It can be seen that with a properly chosen DAUIP information alphabet, the CE-MRT power gap
is roughly $3.5$ dB for a desired information rate of $3$ bpcu.
Also,  the CE-MRT power gap is small even for $N=2$, which makes
 CE transmission possible for  conventional TX with few antennas.

We now investigate the ratio $C_{\mbox{\footnotesize donut}}/C_{\mbox{\footnotesize ATPC}}$ at high $P_T M({\bf h})^2/\sigma^2$.
For $N=1$, it is known that, at large $P_T \vert h_1 \vert^2 /\sigma^2$ (i.e., large $C_{\footnotesize \mbox{ATPC}}$),
capacity with a CE input is
roughly {\em half} of the channel capacity under ATPC \cite{Wyner}.
This fact is illustrated in Fig.~\ref{fig_N116_ce_cap}, where, for $N=1$ the channel capacity under CE transmission
has a much smaller slope w.r.t. $P_T/\sigma^2$ as compared to the slope of the channel capacity under ATPC.
For $N > 1$, using (\ref{unif_rate1}), (\ref{C_MRT}) and (\ref{cap_cmp_eq}) it can be shown that
{
\begin{eqnarray}
\label{rate_compare}
1 > \frac{C_{\mbox{\footnotesize donut}}}{C_{\mbox{\footnotesize ATPC}}} > \frac {I(y ; u)^{\mbox{\footnotesize{unif}}}}{C_{\footnotesize \mbox{ATPC}}} \geq 1 - \frac{\log_2{\big (}\frac{1}{\kappa}{\big )}}{C_{\footnotesize \mbox{ATPC}}} .
\end{eqnarray}
}
For the i.i.d. Rayleigh fading channel and i.i.d. fading channels with bounded channel gains, the convergence in (\ref{k_limit}) implies that, for any arbitrary $\epsilon > 0$, there exists an integer $N(\epsilon)$ such that with $N > N(\epsilon)$, the probability that a
channel realization will have a value of $\kappa \geq \frac {({\mathbb E}[ \vert h_i \vert])^2} { e {\mathbb E}[ \vert h_i \vert^2]} - \epsilon$
is greater
than $1 - \epsilon$.
For single-path DLOS channels we already know that $\kappa = 1/e$ for $N > 1$.
Compared to $N =1$, with $N \gg 1$ and high $P_T  M({\bf h})^2/\sigma^2$, from (\ref{rate_compare}) it follows that CE transmission can
achieve an information rate {\em close} to the capacity $C_{\footnotesize \mbox{ATPC}}$ under ATPC, since $1 - \frac{\log_2(1/\kappa)}{C_{\footnotesize \mbox{ATPC}}}$ is {\em
close to $1$} (as $C_{\footnotesize \mbox{ATPC}}$ is large, and $\kappa$ is greater than a positive constant with high probability), i.e.
{
\vspace{-3mm}
\begin{eqnarray}
\frac{C_{\mbox{\footnotesize donut}}}{C_{\mbox{\footnotesize ATPC}}} \approx 1 \,\,\,\,\,\, \mbox{for} \,\,\, N \gg 1 \,\,\,,\,\,\, \frac{P_T}{\sigma^2} M({\bf h})^2 \gg 1.
\end{eqnarray}
}
This fact is illustrated through Fig.~\ref{fig_N4_ce_cap} and Fig.~\ref{fig_N64_ce_cap}, where it can be seen that
for both $N=4$ and $N=64$, the slope of the ergodic information rate achieved with per-antenna CE transmission (with information symbols
uniformly distributed inside the doughnut set)
is the same as the slope of the ergodic channel capacity under ATPC.
Similar observations can be made from Fig.~\ref{fig_N116_ce_cap} with DAUIP alphabets.
The intuitive reasoning for this observation is as follows.
For $N = 1$, the doughnut set is a circle in the complex plane, due to which information symbols have the same amplitude
and differ from each other only in the phase (i.e., they exploit only one degree of freedom for information transmission). In contrast, with $N > 1$, the doughnut set includes all complex numbers with amplitude
in the range $[m({\bf h}) \,,\, M({\bf h})]$, which implies that information symbols can vary in both phase and amplitude (exploiting
both degrees of freedom).
\vspace{-5mm}
\section{Achievable Array Power Gain}
For a desired rate $R$ and a given precoding scheme, with $N$ antennas, the {\em array power gain} achieved by this scheme is defined to be the factor of reduction in the total transmit power required
to achieve a fixed rate of $R$ bpcu, when the number of TX antennas is increased
from $1$ to $N$.
Under ATPC, with $N$ antennas the MRT precoder
achieves an array power gain of (using (\ref{C_MRT}))
{
\vspace{-2mm}
\begin{eqnarray}
G^{{\footnotesize\mbox{MRT}}}_N(R) = \frac {\sum_{i=1}^N \vert h_i \vert^2 } { \vert h_1 \vert^2}
\end{eqnarray}
}
which is $O(N)$ for i.i.d. fading and DLOS.
With CE transmission, using the R.H.S of (\ref{unif_rate1}) as the achievable
information rate, the array power gain achieved with $N$ antennas is given by
{\small
\begin{eqnarray}
\label{apg}
G^{{\footnotesize\mbox{CE}}}_N(R)
 =  N \,\, \frac{G^{\mbox{CE}}_2(R) }{2} \frac {  {\Big \{} {\big \{} \sum_{i=1}^N \vert h_i \vert / N {\big \} }^2 - m({\bf h})^2/N    {\Big \}} }   { {\Big \{} {\big \{} \sum_{i=1}^2 \vert h_i \vert / 2 {\big \} }^2 - (\vert h_1 \vert - \vert h_2 \vert)^2 / 4   {\Big \}}  }
\end{eqnarray}
}
\normalsize
where $G^{\mbox{CE}}_2(R)$ is the array power gain achieved with only $2$ antennas and depends only on $h_1$ and $h_2$.
From (\ref{apg}), it is clear that
$G^{{\footnotesize\mbox{CE}}}_N(R)$ is $O(N)$ for i.i.d. Rayleigh fading, i.i.d. fading with bounded channel gains and DLOS (for i.i.d.
Rayleigh fading and i.i.d. fading with bounded channel gains, $\sum_i \vert h_i \vert /N \rightarrow_p {\mathbb E}[ \vert h_i \vert]$ and $m({\bf h})^2 / N \rightarrow_p
0$ as $N \rightarrow \infty$).
Therefore, for practical fading scenarios like i.i.d. Rayleigh fading, i.i.d. fading with bounded channel gains and DLOS,
{\em an $O(N)$ array power gain can indeed be achieved even with per-antenna CE transmission}.

\begin{figure}[t]
\begin{center}
\hspace{-4mm}
\epsfig{file=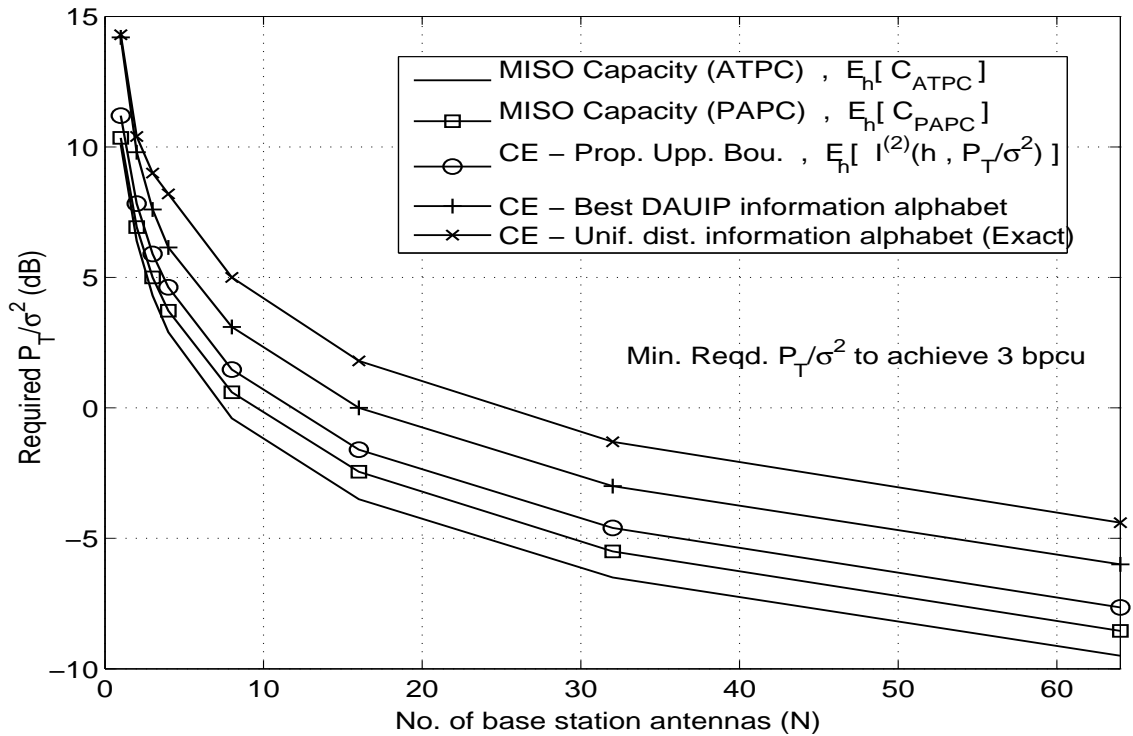, width=160mm,height=105mm}
\end{center}
\vspace{-4mm}
\caption{Minimum $P_T/\sigma^2$ required to achieve an ergodic information rate of $3$ bpcu as a function of  the number of antennas $N$, for i.i.d. ${\mathcal C}{\mathcal N}(0,1)$ Rayleigh fading.}
\label{fig_apg_3bpcu}
\vspace{-4mm}
\end{figure}

This conclusion is validated in Fig.~\ref{fig_apg_3bpcu}, where we plot the minimum $P_T/\sigma^2$ required by the
CE, MRT, and the PAPC precoder to achieve an ergodic information rate of $R=3$ bpcu.
For all precoders, it is observed that, at sufficiently large $N$, the required $P_T/\sigma^2$ reduces by roughly
$3$ dB with every doubling in the number of TX antennas. This confirms the fact that, an $O(N)$ array power gain can be achieved
even with per-antenna CE transmission.
The minimum required $P_T/\sigma^2$ is also tabulated in Table \ref{table_1}.
\section{Outage probability under per-antenna CE transmission}
In scenarios where the channel coherence time
is much longer than the end-to-end delay requirements and where a constant data throughput
rate is desired, we are faced with the possibility of an outage, wherein the
channel capacity is less than the desired information rate. The outage probability under ATPC is defined as
$P_{\footnotesize \mbox{out}}^{\footnotesize \mbox{ATPC}}(R,P_T/\sigma^2) \Define \mbox{Prob}(C_{\footnotesize \mbox{ATPC}} \leq R) = \mbox{Prob}(\Vert {\bf h} \Vert^2 \leq (2^R - 1)\sigma^2/P_T)$ where $R$ is the desired constant information rate.
To have a low outage probability, one needs to
increase the total transmit power $P_T$. With large $N$, due to the increased degrees of freedom
in the r.v.\ $\Vert {\bf h} \Vert^2$ ($\chi^2$ distributed with $2N$ degrees of freedom for i.i.d. Rayleigh distributed channel gains) it is clear that, {\em under ATPC the slope of the outage
probability for the MISO channel w.r.t. $P_T/\sigma^2$ increases with increasing $N$} (on a log-log plot this slope
in the asymptotic limit of $P_T/\sigma^2 \rightarrow \infty$ is commonly known as the ``diversity'' order).
Further, a higher slope at large $N$ implies that less extra $P_T$ would be required to achieve a fixed decrease in the desired outage probability.
However, it is not clear, as to whether the above conclusion is valid even under per-antenna CE 
transmission.

Using the proposed upper and lower bound to $C_{\footnotesize \mbox{donut}}$ (see Sections~\ref{donut_ch_low_bnd_sec} and \ref{donut_ch_upp_bnd_sec}) we can derive
lower and upper bounds
to the outage probability of the proposed CE precoder.
{
The outage probability of the proposed CE precoder is given by
\begin{eqnarray}
\label{lb_bnd1}
P_{\footnotesize \mbox{out}}^{\footnotesize \mbox{CE}}(R,P_T/\sigma^2) & \Define & \mbox{Prob}(C_{\footnotesize \mbox{donut}} \leq R) \nonumber \\
& \geq & \mbox{Prob}{\Big (} I^{(2)} {\Big (} {\bf h},\frac{P_T}{\sigma^2} {\Big )}  \leq R {\Big )}
\end{eqnarray}
where the second inequality follows from the upper bound to $C_{\footnotesize \mbox{donut}}$ in (\ref{Iuy_finalbetabnd3}),
since ${\{}I^{(2)} {\Big (} {\bf h},\frac{P_T}{\sigma^2} {\Big )}  \leq R\}$ implies that ${\{}C_{\footnotesize \mbox{donut}} \leq R\}$.
Similarly, by using the lower bound to $C_{\footnotesize \mbox{donut}}$ in (\ref{unif_rate1}) we get the following
upper bound on $P_{\footnotesize \mbox{out}}^{\footnotesize \mbox{CE}}(R,P_T/\sigma^2)$
\begin{eqnarray}
\label{ub_bnd1}
P_{\footnotesize \mbox{out}}^{\footnotesize \mbox{CE}}(R,P_T/\sigma^2) & \leq & \mbox{Prob}{\Big (}\log_2 {\Big (} 1 + \frac{P_T}{\sigma^2} \frac{M({\bf h})^2 -  m({\bf h})^2}{e} {\Big )}  \,\leq \, R {\Big ) }
\end{eqnarray}
The diversity order achieved is defined as
\begin{eqnarray}
d_{\footnotesize \mbox{out}}^{\footnotesize \mbox{CE}} & \Define & \lim_{\frac{P_T}{\sigma^2} \rightarrow \infty} \frac{- \log (P_{\footnotesize \mbox{out}}^{\footnotesize \mbox{CE}}(R,P_T/\sigma^2))}{\log(P_T/\sigma^2)}
\end{eqnarray}
\emph{In Appendix \ref{app_B} we analytically show that $d_{\footnotesize \mbox{out}}^{\footnotesize \mbox{CE}} \geq (N - 1)$
for the i.i.d. ${\mathcal C}{\mathcal N}(0,1)$ Rayleigh fading channel.}
This result is tight for large $N$, since the maximum achievable diversity order is $N$.
}

\begin{figure}[t]
\begin{center}
\hspace{-4mm}
\epsfig{file=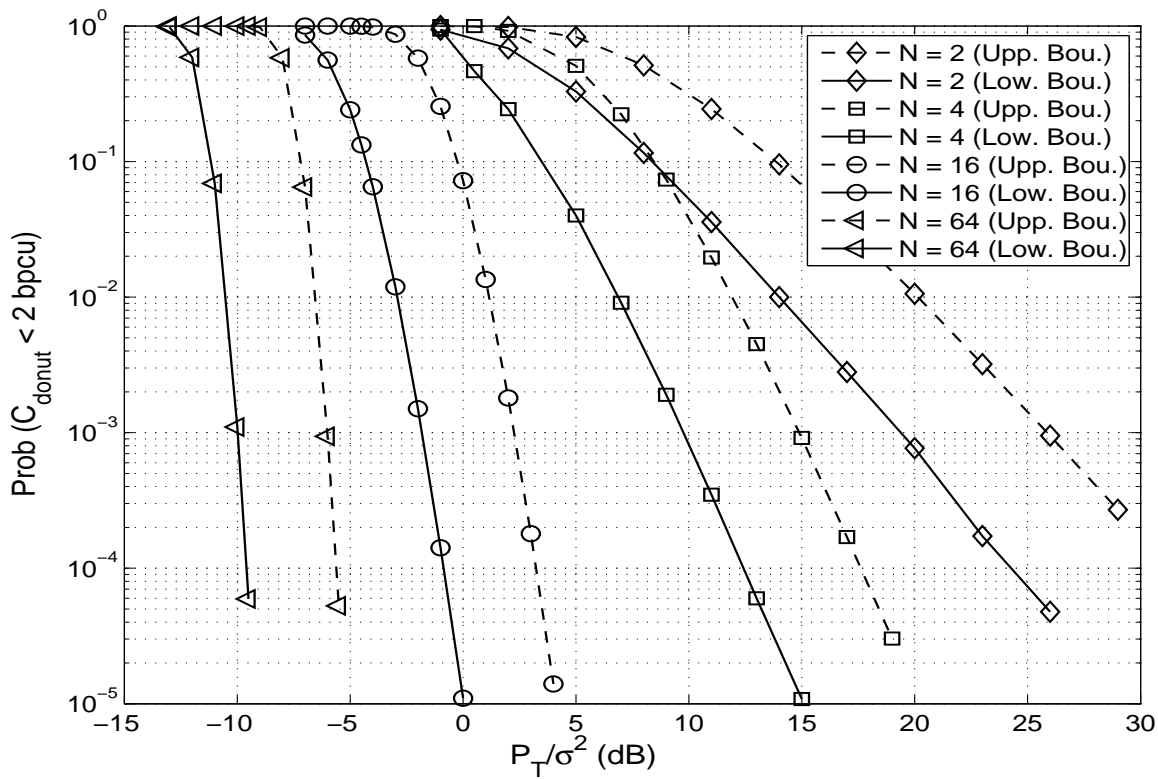, width=155mm,height=105mm}
\end{center}
\vspace{-4mm}
\caption{Upper and lower bounds on the outage probability of the proposed CE precoder at rate 2 bpcu, as a function of
$P_T/\sigma^2$, for i.i.d. ${\mathcal C}{\mathcal N}(0,1)$ Rayleigh fading.}
\label{fig_pout_ce}
\vspace{-4mm}
\end{figure}

We support the above conclusion through Fig.~\ref{fig_pout_ce}, where we plot the upper and lower bounds on the outage probability
of the proposed CE precoder as a function of $P_T/\sigma^2$ for
$N=2,4,16,64$ (i.i.d. Rayleigh fading). The bounds on the right hand side of (\ref{lb_bnd1}) and (\ref{ub_bnd1}),
have been computed through simulations.
It can be seen that for a constant desired rate of $R=2$ bpcu, the slope of the outage probability
curves increase with increasing $N$.
\section{Overall Improvement in Power Efficiency by using CE Transmission}
On one hand, with CE transmission we improve the power efficiency by enabling the use of   highly power-efficient amplifiers,
but at the same time, restricting the per-antenna channel inputs to CE (since
highly power-efficient amplifiers are generally non-linear) requires extra transmit power
(compared to Gaussian inputs) to achieve a fixed desired information rate.
If this extra transmit power is significantly smaller than the improvement in power
efficiency gained by using highly power-efficient amplifiers, then it is clear that 
using per-antenna CE transmission will lead to an overall gain in power efficiency.

Motivated by the above discussion, for a TX with $N$ antennas, compared to using highly linear and power-inefficient
amplifiers with Gaussian inputs (MRT precoder), the overall gain in power efficiency
by using highly power-efficient amplifiers with per-antenna CE inputs is given by
$\rho  \Define  { \frac{\mbox{PAE}_{\mbox{\small non-linear}}} { \mbox{PAE}_{\mbox{\small linear}}}} \, / \,  { P_{\footnotesize \mbox{gap}}^{\footnotesize \mbox{CE},\mbox{MRT}} }$
where $\mbox{PAE}_{\mbox{\small non-linear}}$ and $\mbox{PAE}_{\mbox{\small linear}}$ denote the power-efficiency of non-linear and
linear power amplifiers respectively.\footnote{\footnotesize{For an RF power amplifier, the power efficiency is the ratio
of the total RF power radiated to the total amplifier input power.}} For a highly linear power amplifier, $\mbox{PAE}_{\mbox{\small linear}} \approx 
0.15 - 0.25$, whereas a highly power-efficient but non-linear amplifier has
$\mbox{PAE}_{\mbox{\small non-linear}} \approx 0.7-0.85$ \cite{cripps}.
As an illustrative example, with $\mbox{PAE}_{\mbox{\small linear}} = 0.2 $ and $\mbox{PAE}_{\mbox{\small non-linear}} = 0.8$,
using analytical results on $P_{\footnotesize \mbox{gap}}^{\footnotesize \mbox{CE},\mbox{MRT}}$ (see Section~\ref{inf_rate_comp_sec}), it follows that
in single-path DLOS and i.i.d.\ Rayleigh fading channels
{\em it is indeed beneficial to
use per-antenna CE inputs with highly power-efficient amplifiers}
($P_{\footnotesize \mbox{gap}}^{\footnotesize \mbox{CE},\mbox{MRT}} \leq 4 \ (6 \mbox{dB})$ implies that
$ \rho > 1$).
At practically interesting low to moderate values of $P_T M({\bf h})^2 / \sigma^2 $,
for i.i.d. Rayleigh fading channels $\rho$ varies from
$4.95$ dB (at rates below $1$ bpcu) to $2.5$ dB (at an information rate of $3$ bpcu).
\vspace{-3mm}
\section{Conclusions and Future Work}
In this paper, we derived an achievable rate for a single-user Gaussian MISO channel under the constraint
that the signal transmitted from each antenna has a constant envelope.
We showed that for i.i.d. Rayleigh fading channels, i.i.d. fading channels with bounded channel gains and DLOS channels, even with the stringent per-antenna CE constraint, an $O(N)$ array power gain can be achieved
with $N$ antennas.
Also, compared to the average-only total transmit power constrained channel,
the extra total transmit power required under the CE constraint to achieve a desired rate (i.e., power gap), is shown to be bounded and small.
We conjecture that these results hold true for a much broader class of fading channels, and are not limited to
i.i.d. Rayleigh fading, i.i.d. fading channels with bounded channel gains and DLOS channels.
We are currently extending the results in this paper to the multi-user setting, see \cite{icassp2012}.

\appendices

\section{{On the order of $m({\bf h})$ as $N \rightarrow \infty$}}
\label{app_mh}
{
Before discussing the main result, we make some definitions.
For a random channel vector ${\bf h} = (h_1, h_2, \cdots, h_N)^T$,
let $Z_i \Define \vert h_i \vert^2$. Further, let $Z_{(i)}\,,\,i=1,2,\cdots,N$ be defined
to be the $i$-th smallest value among $Z_1, \cdots, Z_N$. Therefore, we have
$0 \, \leq \, Z_{(1)} \, \leq \, Z_{(2)} \, \leq \, \cdots \, \leq \, Z_{(N)} \, < \, \infty$.
}
{
\begin{mytheorem}\label{mh_o}
For an i.i.d. ${\mathcal C}{\mathcal N}(0,1)$ Rayleigh fading channel, for any constant $c > 0$
\begin{equation}
\label{mh_o_st}
\lim_{N \rightarrow \infty} \mbox{\footnotesize{Prob}}{\Big (} m({\bf h}) \, \geq \, \frac { c \log(N)}{\sqrt{N}}   {\Big )}  \,\, = \,\, 0
\end{equation}
where $m({\bf h})$ has been defined in (\ref{max_min_def}).
\end{mytheorem}
}

{
{\it Proof} --
It suffices to prove that
{
\vspace{-4mm}
\begin{equation}
\label{mh_o_prf1}
\lim_{N \rightarrow \infty} \mbox{\footnotesize{Prob}}{\Big (} m({\bf h}) \, \leq \, \frac { c \log(N)}{\sqrt{N}}   {\Big )}  \,\, = \,\,1 
\end{equation}
}
Further, since $m({\bf h}) \leq \frac{\Vert {\bf h} \Vert_{\infty}}{\sqrt{N}} = \frac {\max_{i=1,\ldots,N} \vert h_i \vert}{ \sqrt{N}}$ (Lemma \ref{min_val_bnd}),
it suffices to show that
\begin{equation}
\label{mh_o_prf2}
\lim_{N \rightarrow \infty} \mbox{\footnotesize{Prob}}{\Big (} \Vert {\bf h} \Vert_{\infty} \, \leq \, { c \log(N)}   {\Big )}  \,\, = \,\,1 
\end{equation}
In terms of the newly defined random variables above, this is equivalent to proving that
\begin{equation}
\label{mh_o_prf3}
\lim_{N \rightarrow \infty} \mbox{\footnotesize{Prob}}{\Big (} Z_{(N)} \, \leq \, { c^2 \log^2N}   {\Big )}  \,\, = \,\,1 
\end{equation}
Due to i.i.d. ${\mathcal C}{\mathcal N}(0,1)$ Rayleigh fading, the random variables $Z_i, i=1,2,\cdots,N$ are i.i.d. exponentially distributed with mean value $1$.
Therefore
{
\vspace{-3mm}
\begin{eqnarray}
\label{mh_o_prf4}
\mbox{\footnotesize{Prob}}{\Big (} Z_{(N)} \, \leq \, { c^2 \log^2N}   {\Big )}  & =  &
\prod_{i=1}^N \mbox{\footnotesize{Prob}}{\Big (} Z_i \, \leq \, { c^2 \log^2N} {\Big )} 
\,\, = \,\, \prod_{i=1}^N {\Big (}  1 \, - \, e^{-c^2 \log^2N} {\Big )} \nonumber \\
&  = & {\Big (}  1 \, - \, e^{-c^2 \log^2N} {\Big )}^N \,\,
 = \,\, {\Big (}  1 \, - \, \frac{1}{N^{c^2 \log N}} {\Big )}^N.
\end{eqnarray}
}
}
{
We next show that
{
\vspace{-4mm}
\begin{eqnarray}
\label{mh_o_prf5}
\lim_{N \rightarrow \infty} \log \mbox{\footnotesize{Prob}}{\Big (} Z_{(N)} \, \leq \, { c^2 \log^2N}   {\Big )} & = & \lim_{N \rightarrow \infty} N \log{\Big (} 1 - \frac{1}{N^{c^2 \log N}} {\Big )}  =  0
\end{eqnarray}
}
from which (\ref{mh_o_prf3}) follows immediately.
To prove (\ref{mh_o_prf5}), note that for any $c > 0$ and all $N > 2$, $N^{c^2 \log N} > 1$.
Further, using the inequality $\log(1 - x) \leq -x$ for $0 \leq x < 1$ \cite{math_hbook}, for $N > 2$ we have
\begin{eqnarray}
\label{mh_o_prf6}
N \log{\Big (} 1 - \frac{1}{N^{c^2 \log N}} {\Big )} & \leq & - \frac{ N} {N^{c^2 \log N}} 
\end{eqnarray}
Using (\ref{mh_o_prf6}) we have
{
\vspace{-4mm}
\begin{eqnarray}
\label{mh_o_prf7}
\lim_{N \rightarrow \infty} N \log{\Big (} 1 - \frac{1}{N^{c^2 \log N}} {\Big )} & \leq & - \lim_{N \rightarrow \infty} \frac{ N} {N^{c^2 \log N}}  = 0.
\end{eqnarray}
}
Using the inequality $\log(1 - x) \geq - x /(1 -x)$ for $0 \leq x < 1$ \cite{math_hbook}, for $N > 2$ we have
\begin{eqnarray}
\label{mh_o_prf8}
N \log{\Big (} 1 - \frac{1}{N^{c^2 \log N}} {\Big )} & \geq & - \frac{ N} {N^{c^2 \log N}}  \frac{1}{1 \, - \, e^{-c^2 \log^2N} }
\end{eqnarray}
which implies that
{
\vspace{-4mm}
\begin{eqnarray}
\label{mh_o_prf9}
\lim_{N \rightarrow \infty} N \log{\Big (} 1 - \frac{1}{N^{c^2 \log N}} {\Big )} & \geq & - \lim_{N \rightarrow \infty} \frac{ N} {N^{c^2 \log N}}  \frac{1}{1 \, - \, e^{-c^2 \log^2N} }  = 0.
\end{eqnarray}
}
Combining (\ref{mh_o_prf9}) and (\ref{mh_o_prf7}) proves (\ref{mh_o_prf5}) which completes the proof.
}
\section{{Diversity analysis for the Outage probability of the proposed CE precoder}}
\label{app_B}
{
Using the lower bound on $C_{\mbox{\footnotesize donut}}$ in (\ref{unif_rate2}), an upper bound on the
outage probability is given by
\begin{eqnarray}
\label{app2_eq1}
P_{\footnotesize \mbox{out}}^{\footnotesize \mbox{CE}}(R,P_T/\sigma^2) & = & \mbox{Prob}(C_{\footnotesize \mbox{donut}} \leq R) \,\,
 \leq \,\, \mbox{{Prob}}{\Big (}\log_2 {\Big (} 1 + \frac{P_T}{\sigma^2}   \frac{ \Vert {\bf h} \Vert_1^2 - \Vert {\bf h} \Vert_{\infty}^2} {N e} {\Big )} \leq R{\Big )}
\end{eqnarray}
In terms of the new random variables defined at the beginning of Appendix \ref{app_mh}, we have
\begin{eqnarray}
\Vert {\bf h} \Vert_1^2 - \Vert {\bf h} \Vert_{\infty}^2 & = & {\Big (} \sum_{i=1}^N \sqrt{Z_{(i)}}  {\Big )}^2  \, - \, Z_{(N)} \,\,
 \geq \,\, \sum_{i=1}^{N-1} Z_{(i)}.
\end{eqnarray}
Using this fact in (\ref{app2_eq1}), we have
\begin{eqnarray}
\label{app2_eq2}
P_{\footnotesize \mbox{out}}^{\footnotesize \mbox{CE}}(R,P_T/\sigma^2)  & \leq & \mbox{{Prob}}{\Big (}\log_2 {\Big (} 1 + \frac{P_T}{\sigma^2}   \frac{ \sum_{i=1}^{N-1} Z_{(i)} } {N e} {\Big )} \leq R{\Big )}
\end{eqnarray}
Let us define random variables
{
\vspace{-4mm}
\begin{eqnarray}
Y_i & \Define & (N - i + 1) \, ( Z_{(i)} \, - \,  Z_{(i-1)} ) \,\,\,\, i=1,2,\cdots,N.
\end{eqnarray}
}
Note that $Y_1 \Define N Z_{(1)}$.
For the i.i.d. ${\mathcal C}{\mathcal N}(0,1)$ Rayleigh fading channel, it is known that $Y_i \in [0 \,,\, \infty)\,,\,i=1,2,\cdots,N$
are i.i.d. exponentially distributed random variables with mean $1$ (see section 2.7, page $17$ in \cite{David}).
From the definition above, it immediately follows that
\begin{eqnarray}
\sum_{i=1}^{N - 1} Y_i & = & Z_{(N -1) } \, + \, \sum_{i=1}^{N - 1} Z_{(i)}
\end{eqnarray}
which implies that
{
\vspace{-6mm}
\begin{eqnarray}
\label{app2_eq3}
\sum_{i=1}^{N - 1} Z_{(i)} & \geq &  \frac{1}{2} \sum_{i=1}^{N - 1} Y_i
\end{eqnarray}
}
since $Y_i$ and $Z_{(i)}$ are non-negative random variables.
Using (\ref{app2_eq3}) in (\ref{app2_eq2}) we have
\begin{eqnarray}
\label{app2_eq4}
P_{\footnotesize \mbox{out}}^{\footnotesize \mbox{CE}}(R,P_T/\sigma^2)  & \leq &  \mbox{{Prob}}{\Big (}\log_2 {\Big (} 1 + \frac{P_T}{\sigma^2}   \frac{ \sum_{i=1}^{N-1} Y_i } {2 N e} {\Big )} \leq R{\Big )} \,\,
 = \,\, \mbox{{Prob}}{\Big (} \sum_{i=1}^{N-1} Y_i \, \leq \, \frac{2 e N (2^R - 1)}{P_T/\sigma^2} {\Big )}
\end{eqnarray}
Since, the event $ \sum_{i=1}^{N-1} Y_i \, \leq \, \frac{2 e N (2^R - 1)}{P_T/\sigma^2} $  implies that each $Y_i \, \leq  \,  \frac{2 e N (2^R - 1)}{P_T/\sigma^2}$,
we further have
\begin{eqnarray}
\label{app2_eq5}
P_{\footnotesize \mbox{out}}^{\footnotesize \mbox{CE}}(R,P_T/\sigma^2)  & \leq &  \mbox{{Prob}}{\Big (}  Y_i \, \leq \, \frac{2 e N (2^R - 1)}{P_T/\sigma^2} \,\,,\,\,i=1,2,\cdots,N-1 {\Big )}
\end{eqnarray}
Since $Y_i$ are i.i.d. exponentially distributed, the right hand side in the above can be further simplified to
\begin{eqnarray}
\label{app2_eq55}
P_{\footnotesize \mbox{out}}^{\footnotesize \mbox{CE}}(R,P_T/\sigma^2)  & \leq &  \prod_{i=1}^{N -1} 
\mbox{{Prob}}{\Big (}   Y_i \, \leq \, \frac{2 e N (2^R - 1)}{P_T/\sigma^2} {\Big )} \,\,
 = \,\, {\Big (} 1 \, - \, e^{- \frac{2 e N (2^R - 1)}{P_T/\sigma^2}} {\Big )}^{N - 1} 
\end{eqnarray}
The diversity order achieved by the outage probability therefore satisfies
\begin{eqnarray}
d_{\footnotesize \mbox{out}}^{\footnotesize \mbox{CE}} & \Define & \lim_{\frac{P_T}{\sigma^2} \rightarrow \infty} \frac{- \log (P_{\footnotesize \mbox{out}}^{\footnotesize \mbox{CE}}(R,P_T/\sigma^2))}{\log(P_T/\sigma^2)} \,\,
 \geq \,\, (N - 1) \lim_{\frac{P_T}{\sigma^2} \rightarrow \infty}  \frac{ - \log{\Big (} 1 \, - \, e^{- \frac{2 e N (2^R - 1)}{P_T/\sigma^2}} {\Big )}} {\log{\Big (} \frac{P_T}{\sigma^2}   {\Big )}}
\end{eqnarray}
where we have used (\ref{app2_eq55}) for the inequality.
Using the identity
\begin{eqnarray}
\lim_{x \rightarrow 0} \frac{ \log(1 - e^{-c x})}{\log x} & = & 1 \,\,\, (c > 0)
\end{eqnarray}
with $x = \sigma^2/P_T$ and $c = 2 e N (2^R - 1) \, > 0$, we have
\begin{eqnarray}
\lim_{\frac{P_T}{\sigma^2} \rightarrow \infty}  \frac{ - \log{\Big (} 1 \, - \, e^{- \frac{2 e N (2^R - 1)}{P_T/\sigma^2}} {\Big )}} {\log{\Big (} \frac{P_T}{\sigma^2}   {\Big )}} & = & 1
\end{eqnarray}
which then proves that
\begin{eqnarray}
d_{\footnotesize \mbox{out}}^{\footnotesize \mbox{CE}} & \geq & (N - 1) 
\end{eqnarray}
}
\newpage
{
\small
\begin{table}
\caption{\normalsize Capacity ratios/power gap of   CE transmission w.r.t. MRT and PAPC transmission}
\small
\centering
\begin{tabular}{| l || l | l | l | l || l | l |}
\hline
& \multicolumn{4}{ l |}{ \normalsize $N \gg 1$  } & \multicolumn{2}{ l |}{ \normalsize $N = 1$  } \\ \hline  
& \multicolumn{2} { l |} { $\frac{P_T}{\sigma^2} M({\bf h})^2 \ll 1 $  } &  \multicolumn{2} { l |} { $\frac{P_T}{\sigma^2} M({\bf h})^2 \gg 1 $  } 
& $\frac{P_T}{\sigma^2} \vert h_1 \vert^2 \ll 1$ & $\frac{P_T}{\sigma^2} \vert h_1 \vert^2 \gg 1$ \\ \hline
& \mbox{\small i.i.d. Rayleigh fading,} & \mbox{\small DLOS} & \mbox{\small i.i.d. Rayleigh fading,} & \mbox{\small DLOS} &  & \\
& \mbox{\small i.i.d. fading channels} &  & \mbox{\small i.i.d. fading channels} &  &  & \\
& \mbox{\small with bounded channel gains} &  & \mbox{\small with bounded channel gains} &  &  & \\\hline
$P_{\footnotesize \mbox{gap}}^{\footnotesize \mbox{CE},\mbox{MRT}} $ & \normalsize ${10 \log_{10}(\frac {  {\mathbb E}[ \vert h_i \vert^2]} {({\mathbb E}[ \vert h_i \vert])^2})}$ & 0 & $ \geq  3 + 10 \log_{10}(\frac {  {\mathbb E}[ \vert h_i \vert^2]} {({\mathbb E}[ \vert h_i \vert])^2})$ &  $\geq 3$ & $0$ & $\infty$ \\   
(dB) &  & & $\leq 4.3 + 10 \log_{10}(\frac {  {\mathbb E}[ \vert h_i \vert^2]} {({\mathbb E}[ \vert h_i \vert])^2})$ & $\leq 4.3 $ & & \\ \hline
$P_{\footnotesize \mbox{gap}}^{\footnotesize \mbox{CE},\mbox{PAPC}} $ & 0 & 0 & $\geq 3$ & $\geq 3$ & $0$ & $\infty$ \\ 
(dB) & & & $\leq 4.3$ & $\leq 4.3$ &  & \\ \hline
$\frac {C_{\mbox{\footnotesize donut}}} {C_{\footnotesize \mbox{ATPC}}}$ &
$ \frac {({\mathbb E}[ \vert h_i \vert])^2}  {  {\mathbb E}[ \vert h_i \vert^2]} $ & 1 & 1 & 1 & 1 & $\frac {1} {2}$ \\ \hline
$\frac {C_{\mbox{\footnotesize donut}}} {C_{\footnotesize \mbox{PAPC}}}$ & 
1 & 1 & 1 & 1 & 1 & $\frac {1} {2}$ \\ \hline
\end{tabular}
\label{table_0}
\end{table}
}
{
\begin{table}
\normalsize
\caption{\normalsize Signal-to-noise-ratio $P_T/\sigma^2$ (dB) required to achieve an ergodic rate of $3$ bpcu 
(i.i.d. ${\mathcal C}{\mathcal N}(0,1)$ Rayleigh fading) }
\centering
\begin{tabular}{| c ||  c |  c | c | c | c | c | c | c |}
\hline
 & N=1 & N=2 & N=3 & N=4 & N=8 & N=16 & N = 32 & N = 64\\
\hline
MRT (ATPC) & 10.2 & 6.4 & 4.3  & 2.9 & -0.4 & -3.5 & -6.5 & -9.5\\
\hline
PAPC & 10.2 & 6.9 & 5.0  & 3.7 & 0.6 & -2.5 & -5.5 & -8.6 \\
\hline
CE (best DAUIP)  & 14.3 & 9.8 & 7.6  & 6.2 & 3.1 & 0 & -3.0 & -6.0 \\
\hline
CE (UNIF)  & 14.3 & 10.4 & 9.0  & 8.2 & 5.0 & 1.8 & -1.3 & -4.4 \\
\hline
\end{tabular}
\label{table_1}
\end{table}
}
\normalsize


\end{document}